\documentclass[aps,prd,showpacs,nofootinbib,twocolumn,superscriptaddress,floatfix]{revtex4}

\usepackage{amsmath,amssymb}
\usepackage[dvips, pdftex]{graphicx}
\usepackage[tight,footnotesize]{subfigure}
\usepackage[usenames]{color}
\usepackage[dvipsnames]{xcolor}
\usepackage[utf8]{inputenc}

\DeclareGraphicsExtensions{.jpg,.mps,.pdf,.png,.eps} 

\newcommand{\be}{\begin{equation}}
\newcommand{\ee}{\end{equation}}
\newcommand{\bea}{\begin{eqnarray}}
\newcommand{\eea}{\end{eqnarray}}
\newcommand{\non}{\nonumber}

\newcommand{\rr}{\mathrm}
\newcommand{\mA}{m_{\rr A}}
\newcommand{\RA}{R_{\rr A}}
\newcommand{\RL}{L}
\newcommand{\WT}{R_{\rr w}}
\newcommand{\rhoW}{\rho_{\rr w}}
\newcommand{\rhoV}{\rho_{\rr v}}
\newcommand{\rhoATM}{\rho_{\rr {atm}}}
\newcommand{\rhoA}{\rho_{\rr {A}}}

\newcommand{\Veff}{V_{\rr {eff}}}
\newcommand{\phibg}{\phi_{\rr {bg}}}
\newcommand{\mbg}{m_{\rr {bg}}}
\newcommand{\GeV}{\ \rr {GeV}}
\newcommand{\dd}{\mathrm{d}}
\newcommand{\mpl}{m_\rr{pl}}

\begin{document}
\title{Probing Modified Gravity with Atom-Interferometry: a Numerical Approach}

\author{Sandrine Schl\"ogel}
\email{sandrine.schlogel@unamur.be}
\affiliation{Namur Center of Complex Systems (naXys), Department of Mathematics, University of Namur, Rempart de la Vierge 8, 5000 Namur, Belgium}
\affiliation{Centre for Cosmology, Particle Physics and Phenomenology,
Institute of Mathematics and Physics, Louvain University,
Chemin du Cyclotron 2, 1348 Louvain-la-Neuve, Belgium}

\author{S\'ebastien Clesse}
\email{clesse@physik.rwth-aachen.de}
\affiliation{Institute for Theoretical Particle Physics and Cosmology (TTK), RWTH Aachen University, D-52056 Aachen, Germany}
\affiliation{Namur Center of Complex Systems (naXys), Department of Mathematics, University of Namur, Rempart de la Vierge 8, 5000 Namur, Belgium}

\author{Andr\'e F\"uzfa}
\email{andre.fuzfa@unamur.be}
\affiliation{Namur Center of Complex Systems (naXys), Department of Mathematics, University of Namur, Rempart de la Vierge 8, 5000 Namur, Belgium}
\affiliation{Centre for Cosmology, Particle Physics and Phenomenology,
Institute of Mathematics and Physics, Louvain University,
Chemin du Cyclotron 2, 1348 Louvain-la-Neuve, Belgium}

\date{\today}

\begin{abstract}
Refined constraints on chameleon theories are calculated for atom-interferometry experiments, using a numerical approach 
consisting in solving for a four-region model the static and spherically symmetric Klein-Gordon equation for the 
chameleon field.   By modeling not only the test mass and the vacuum chamber but also its walls and the exterior 
environment, the method allows to probe new effects on the scalar field profile and the induced acceleration of atoms.  
In the case of a weakly perturbing test mass, the effect of the wall is to enhance the field profile and 
to lower the acceleration inside the chamber by up to one order of magnitude. In the thin-shell regime, 
results are found to be in good agreement with the analytical estimations, when 
measurements are realized in the immediate vicinity of the test mass.   Close to the vacuum chamber wall, the 
acceleration becomes negative and potentially measurable.  This prediction could be used to discriminate between 
fifth-force effects and systematic experimental uncertainties, by doing the experiment at several key positions inside 
the vacuum chamber. For the chameleon potential $V(\phi) = \Lambda^{4+\alpha} / \phi^\alpha$ and a coupling function 
$A(\phi) = \exp(\phi /M)$, one finds $M \gtrsim 7 \times 10^{16} \GeV$, independently of 
the power-law index.  For $V(\phi) = \Lambda^4 (1+ \Lambda/ \phi)$, one finds $M \gtrsim 
10^{14} \GeV$. A sensitivity of $a\sim 10^{-11}  \rr{m/s^2} $ would probe the model up to the Planck scale.
Finally, a proposal for a second experimental set-up, in a vacuum room, is presented. 
In this case, Planckian values of $M$ could be probed provided that $a \sim 10^{-10} \rr{m/s^2}$, a limit reachable
by future experiments. Our method can easily be extended to constrain other models with a screening 
mechanism, such as symmetron, dilaton and f(R) theories.  

\end{abstract}

\maketitle

\section{Introduction}\label{sec1}

The accelerated cosmic expansion, highlighted by several observations such as type-Ia supernovae \cite{Riess}, the 
temperature 
fluctuations of the cosmic microwave background \cite{Planck2015} and the distribution of large scale structures, has 
been the 
subject 
of intense research. The simplest explanation invoking a cosmological constant, in fair agreement with 
current 
observations, suffers from some fine-tuning and coincidence issues.  A wide range of alternative models 
from various frameworks have been proposed (see e.g.~\cite{Amendola} for a review), a 
lot of them introducing a dynamical scalar 
field coupled to matter. In order to leave observational signatures on the structure formation or the expansion history, while passing tight constraints coming from 
laboratory experiments \cite{Adelberger, Upadhye, Kapner}, tests in the solar system (see \cite{Will} for a review, 
also 
\cite{CAST, Rees}) and on galactic scales \cite{Pourhasan, Jain}, a screening mechanism appears to be 
fruitful by suppressing the fifth-force induced by the coupled scalar field in local environments.  
Chameleons \cite{KhouryWeltmanPRL, KhouryWeltmanPRD,Gubser, Brax, Mota, MotaShaw} 
are typical models where the scalar field is suppressed in a 
dense medium,
like in the Solar system, 
while 
acting as a cosmological constant on sparse environment like in the cosmos at late time.  

Many experimental tests of the chameleons have been proposed so far, involving e.g. ultra-cold neutrons \cite{Jenke, 
Lemmel, Brax_neutron, Brax_neutron2, Ivanov, BraxBurrage} or cold atoms interferometry \cite{Fixler, Lamporesi} (see 
also \cite{Shih1, Shih2, Haroche,Sukenik, Baum, Harber, Kasevich, 
Cronin, BraxCasimir, Harber}), still leaving a part of the parameter space unconstrained.
Very recently, new experiments based on atom interferometry, involving a 
source mass inside a vacuum chamber, have been 
proposed to test chameleon models with a high sensitivity \cite{Burrage, khoury}, individual atoms being 
sufficiently small to let the 
scalar field unscreened even if the nucleus is dense.  The experimental setup consists in measuring 
or constraining the 
additional acceleration on individual atoms, due 
to the scalar field gradient induced by the presence of a source mass at the center of the chamber. Forecasts were provided in 
\cite{Burrage} and first experimental results of~\cite{khoury} claimed to rule out most of the chameleon parameter space.

Nevertheless those results rely on analytical assumptions and have not been validated when accounting for the entire 
environment surrounding the experiment.  
In this paper, we consider a four-region model, including the wall of the chamber and the exterior environment, and we solve numerically the Klein-Gordon (KG) equation in the 
static 
and spherically symmetric case. The only assumptions made here are (1) that the scalar field reaches its attractor (the field value at the minimum of its effective potential) 
outside the vacuum chamber, i.e. in the air, at 
spatial infinity,
and (2) that we have a non-singular  
$C^1$ solution everywhere.
Our method reveals that in the regime where the test mass perturbs weakly the scalar field profile, its 
amplitude is controlled by the attractor outside the chamber instead of being related to the chamber size.   
In addition, our analysis presents the advantage of including the important
effects of the vacuum chamber wall that have been neglected so far. We show that the scalar field profile and 
the resulting acceleration can differ by up to one order of magnitude in the weakly perturbing regime, compared to 
previous analysis \cite{Burrage, khoury}. Thus the effects of the wall cannot be neglected in a precise investigation of 
the chameleon parameter space. In the thin-shell regime, analytical approximations are validated to a good accuracy.  
Nevertheless our method highlights order of ten percents deviations.  It also allows to determine quantitatively the 
negative acceleration close to the walls of the vacuum chamber.

We also propose a new experimental set-up, where the atom interferometer is placed inside a 
vacuum room.
Refined constraints on the chameleon parameters (essentially the coupling 
function) are provided for both experimental set-ups and the experimental requirements to exclude the chameleon 
parameter space up to the Planck scale in the future are evaluated. 

The numerical analysis has been performed for two typical chameleon potentials, with varying 
power-laws.  The first one (referred as Chameleon-1), $V(\phi) = \Lambda^{4+\alpha}/ \phi^\alpha$~\cite{Ratra:1987rm} is 
able to reproduce the cosmic acceleration and to fit supernovae data, however tests of general relativity in the solar 
system would exclude the corresponding parameter space~\cite{Hees}.  It is nevertheless interesting to consider this 
model as an illustrative example of field configurations inside the vacuum chamber that are weakly perturbed by the 
presence of the central source mass.   Laboratory experiments also probe other regions of its parameter space, even if 
they are not of direct interest for cosmology.   The second considered potential (referred as Chameleon-2) has an 
additional cosmological constant~\cite{Brax:2004qh},  $V(\phi) = \Lambda^4 \exp\left(\overline{\Lambda}^{\alpha}/ \phi^\alpha\right)\simeq \Lambda^4 (1+\overline{\Lambda}^{\alpha}/ \phi^\alpha)$ where we assume $\Lambda=\overline{\Lambda}$ for keeping only one additional parameter, $\alpha$ being fixed.  For both models we take an 
exponential coupling function $A(\phi) = \exp(\phi / M)$.  


The paper is organized as follow. In Sec.~\ref{sec2}, we briefly remind the equations of motion and introduce 
the considered models.  The experimental setup is described in Sec.~\ref{sec3} and the numerical strategy is detailed in Sec.~\ref{sec4}.  
Analytical results are reminded in Sec.~\ref{sec5} and are compared to the numerical results in Sec.~\ref{secChamelNum} for the two considered models, refined constraints being established on their parameters.  
We finally discuss our results and draw some conclusions and perspectives in Sec.~\ref{sec_CCL}.

\section{The models}\label{sec2}

We start from a generic action for modified gravity models involving a non-minimally coupled dynamical scalar field $\phi$, written
in the Einstein frame,
\bea
S=\int \dd^4x \sqrt{-g} \left[\frac{R}{2\kappa}-\frac{1}{2} \left(\partial\phi\right)^2-V(\phi)\right]\non
\\
+ S_m\left[A^2\left(\phi\right)g_{\mu\nu};\psi_\rr{m}\right],
\eea
with $R$, the scalar curvature, $\kappa=8\pi/\mpl^2$, $\mpl$ being the Planck mass, $\psi_\mathrm{m}$ the 
matter 
fields, $V(\phi)$ a general potential and $A(\phi)$ a general coupling function.

Varying this action with respect to $\phi$ gives the KG equation
\bea
\Box\phi={{\dd V}\over{\dd \phi}}-T{{\dd \ln A}\over{\dd \phi}},
\eea
where $T$ is the trace of the energy-momentum tensor $T_{\mu\nu}=-(2/\sqrt{-g})\left(\partial 
S_\rr{m}/\partial 
g^{\mu\nu}\right)$. We introduce $\tilde{T}_{\mu \nu}$ the stress-energy
tensor for a perfect fluid in the Jordan frame in order to consider conserved quantities, i.e. $\tilde{\nabla}_\mu 
\tilde{T}^{\mu\nu}=0$, the tilde denoting Jordan frame quantities. Energy density $\rho$ and pressure $p$ in 
both frames are then related by~\cite{Damour:1993id}
\bea
\rho&=&A^4\left(\phi\right)\tilde{\rho},
\label{rho}
\\
p&=&A^4\left(\phi\right)\tilde{p},
\eea
so that in the weak field regime the KG equation in the static and spherically symmetric case becomes
\bea  \label{eq:KG}
\phi''+\frac{2}{r} \phi'={{\dd V_{\rm eff}}\over{\dd \phi}}, \hspace{1cm} {{\dd V_{\rm 
eff}}\over{\dd \phi}}={{\dd V}\over{\dd \phi}} + 
\tilde{\rho} A^3 {{\dd A}\over{\dd \phi}},
\eea
in which we have introduced an effective potential $V_{\rm eff}$, the prime denoting a radial coordinate derivative. 
We work here 
in the non-relativistic limit with negligible 
metric potentials and pressure. In Table~\ref{tab:models}, the two considered inverse power-law chameleon potentials and the coupling function $A(\phi)$ are specified. 
Equations of motion for both models are identical since only the derivative 
of the potential $\rr{d}V/\rr{d}\phi$ 
contributes to the KG equation. However viable $\Lambda$ values 
are different.  
For the Chameleon-1 model, $\Lambda$ must be fixed by the supernovae best-fit and obeys to~\cite{Schimd},
\be
\log \Lambda \rr{[GeV]} \approx\frac{19\alpha-47}{4+\alpha},
\ee
in order to reproduce the acceleration of the Universe expansion
even in the presence of the non-minimal coupling \cite{Hees}. In the case of Chameleon-2, $\Lambda$ is
the cosmological constant value $\Lambda \simeq 2.4\, \rm{meV}$ while $\alpha$ is an independent 
variable.


\begin{table*} 
\begin{tabular}{|c|c|c|c|c|c|} 
\hline
 & Potential $V(\phi)$  &  Coupling function $A(\phi)$ & Parameters & Minimum of $V_\rr{eff}$ 
$\left(\phi_\rr{min}\right)$ & Mass sq. at minimum ($m_{\rr{min}}^2$)  \\
\hline
Chameleon 1 & $\Lambda^{4+\alpha}/\phi^\alpha$ & ${\rm{e}}^{\phi/M}$ & $\left(\alpha,\Lambda\right), \, M$ & 
$\left(\frac{\alpha \Lambda^{\alpha+4}M}{\tilde{\rho}}\right)^{1/(\alpha+1)}$ & $\alpha (1+ \alpha) \Lambda^{4+\alpha}  \left( \frac{\tilde \rho}{\alpha M \Lambda^{4+\alpha}} \right)^{\frac{2+\alpha}{1+\alpha}}$ \\
\hline
Chameleon 2  & $ \Lambda^4\left(1+\Lambda^{\alpha}/\phi^\alpha\right) $ & 
${\rm{e}}^{\phi/M}$ & 
$\alpha,\Lambda, M$ & $\left(\frac{\alpha \Lambda^{\alpha+4} 
M}{\tilde{\rho}}\right)^{1/(\alpha+1)}$  & $\alpha (1+ \alpha) \Lambda^{4+\alpha}  \left( \frac{\tilde \rho}{\alpha M \Lambda^{4+\alpha}} \right)^{\frac{2+\alpha}{1+\alpha}}$ \\
\hline
\end{tabular}
\caption{Characterization of the effective potential for the models of interest: potential 
$V(\phi)$, coupling functions 
$A(\phi)$, the model parameters, the minimum
of the effective potential and the corresponding mass. Dependent parameters appear into brackets. Values in the last two columns are valid as 
long as $A(\phi)\simeq1$ (an assumption no longer valid for some models, see discussion in
Sec.\ref{secChamelNum}). } 
\label{tab:models}
\end{table*}

\section{Experimental setup}\label{sec3}

Laboratory experiments measuring the acceleration induced by a test mass can be used to probe and constrain 
modifications of gravity.  
As a reminder, the chameleon field is screened in high density environments while it mediates long-range force in 
sparse ones.
Therefore atomic particles in a ultra-high vacuum chamber can mimic cosmos conditions. In the first experiments, the 
test mass was located outside the vacuum chamber \cite{Fixler, 
Lamporesi}, an 
experimental setup which is not ideal given that the chamber wall screens the fifth force on the atoms. New experiments 
have been proposed in 
\cite{Burrage, khoury} where the
test mass is located inside the vacuum chamber, which improves the constraints on the acceleration due to the scalar 
field. 
Here we focus on a recently proposed atom interferometry experiment \cite{khoury} 
where one takes advantage of matter-wave properties of cesium-133 atoms
in a Fabry-Perot cavity. When an atom absorbs/emits a photon, it recoils with a momentum $p=\hbar k$, with $k$ the 
wavenumber of the absorbed/emitted photon. So, 
one can reproduce the equivalent of a Max-Zehnder interferometer for cold atoms
with three light pulses using counter-propagating laser beams. 
Atoms are initially prepared in a hyperfine state $F=3$ and stored in a 2 dimensional magneto-optical trap.
A first light pulse splits the matter-wave packet in two hyperfine state $F=3$ and $F=4$ and gives an impulse of $\hbar 
k_{\rm eff}$ to the atoms. The effective wavenumber $k_{\rm eff}$ depends on the two counterpropagating beam 
wavenumbers. The probability of hyperfine transition can be 
controlled by the intensity and duration of both 
laser beams.
The second pulse reverses the relative motion of the beams like the mirror of Max-Zehnder interferometer and  
the third pulse acts like a beam splitter which allows overlap of partial matter wave packets. 
Because of the recoil of the atoms, the phase difference between the two arms of the interferometer $\Delta \phi$ is 
a function of the acceleration $a$ of atoms,
\be
\Delta \phi=k_{\rm eff} a T^2,
\ee
where $T\sim 10 \,\rm{ms}$ in general, is the time interval between two pulses. 
To alleviate some systematics effects, counterpropagating laser beams are reversed 
and the aluminum sphere can be positioned in two 
places: a \textit{near} and a \textit{far} positions (the test mass surface is respectively located 8.8 mm and 3 cm far from the atoms), 
which allows to disentangle the 
contribution from chameleon force to Earth's 
gravity.
One measurement consists thus of four interference fringes, corresponding to reversed counterpropagating laser beams 
and both positions of the test mass.
Using this setup, the acceleration induced by the chameleon has been excluded up to
\be
a_\rr{exp}<5.5 \,\rm{\mu m}/s^2 \hspace{1cm}\text{at 95$\%$ C.L.}.
\label{eq:acc_exp}
\ee
The experimental setup proposed in \cite{Burrage} is similar, except that they plan to use cooled rubidium atoms 
launched in a small fountain located 1 cm far from the test mass.
Our numerical simulations can be easily adapted for such a configuration. 

\begin{figure}
\begin{center}
\includegraphics[scale=0.35]{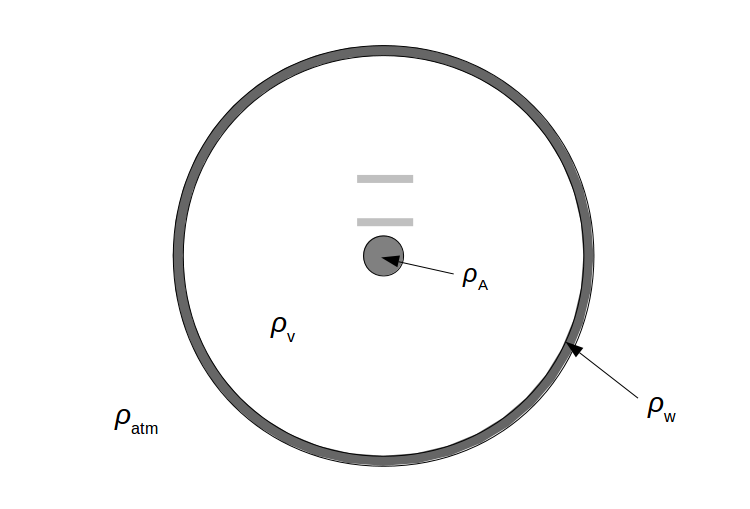}
\caption{Outline of the atom-interferometry experiment, simulated by a four-region model including the source mass, the vacuum chamber, its walls and the exterior environment. In light gray, the \textit{near} and \textit{far} 
positions where the acceleration on atoms is measured (note that we consider a fixed source mass to keep spherical symmetry whereas in the real experimental setup the source mass is moved \cite{khoury}).}
\centering
\label{plot_exp}
\end{center}
\end{figure}

Details of the considered experimental setup 
are reported in Table~\ref{tab:expconf}. 
The size and density of the central mass, the geometry of the chamber and the vacuum density are those of \cite{khoury, Burrage}.  
In addition we consider the thickness and density of the vacuum chamber walls, as well as the exterior density.
In Fig.\ref{plot_exp}, we draw the experimental setup considered
in our numerical simulations. The four regions are labeled by their densities\footnote{In the following of the paper, 
$\rho$ refers to the density in the Jordan frame.}: (1) 
the test mass made of aluminum (${\rho}_A$), (2) the vacuum where the acceleration due to the chameleon is measured 
(${\rho}_{\rm v}$), (3) the wall of the chamber (${\rho}_{\rm w}$) made of stainless steel, (4) the exterior of the chamber, mostly filled by air 
at atmospheric pressure (${\rho}_{\rm 
atm}$). 

\begin{table} 
\begin{tabular}{|c|c|c|} 
\hline
$\RA$ & Radius of the test mass  &  $1 \rm{cm}  / 5.1\times10^{13} \rm{GeV}^{-1} $  \\
$\RL$ & Radius of the chamber & $10 \rm{cm}  / 5.1\times10^{14} \rm{GeV}^{-1} $  \\
$\WT$ & Wall thickness & $1 \rm{cm}  / 5.1\times10^{13} \rm{GeV}^{-1} $   \\
$\mA$ & Test mass  &  11.3g /$6.7\times 10^{24}\rm{GeV}$  \\
$\rhoA$ & Test mass density  & $1.2\times 10^{-17}\rm{GeV}^4$ \\
$\rhoW$ & Wall density &  $3.5 \times 10^{-17} \rm{GeV}^4$ \\
$\rhoV$ & Vacuum density & $5.0 \times 10^{-35} \rm GeV^4$ \\
$\rhoATM $ & Air density $\left(P_\rr{atm}\right)$ & $ 5.2\times 10^{-21} \rm GeV^4$ \\
\hline
\end{tabular}
\caption{Fiducial experimental parameters, corresponding to the 
setup of \cite{khoury}.}
\label{tab:expconf}
\end{table}

\section{Numerical strategy}\label{sec4}

Analytical approaches have been considered so far \cite{Burrage, 
khoury}, which are 
valid under some assumptions like negligible chamber 
wall effects. 
Therefore, numerical methods are useful to validate and refine analytical results, 
by including the effects due to the experimental setup, like the 
thickness and the density of the wall as well as the exterior environment.   In the future, numerical results will be also helpful to study 
more realistic situations where the vacuum chamber is not exactly spherical or cylindrical.

We consider two methods for solving the KG equation \eqref{eq:KG}: a singular and multipoint boundary
value problem (bvp) solver with unknown parameter 
and a non-linear bvp solver implementing up to sixth order a mono-implicit Runge-Kutta method 
with an adaptative mesh refinement, working in \texttt{quad} precision\footnote{For this purpose we have used the 
Matlab function \texttt{bvp4c} which deals with singular bvp's and a modified version of the \texttt{mirkdc} 
bvp solver with adaptative mesh in Fortran.}.
In the latter case, the density in the four regions was made continuous by considering arctan profiles with negligible widths. 

We take the minimal assumption which states that the scalar field is settled to its 
attractor at spatial infinity, i.e. 
$\phi_\infty=\phi_{\rm min}(\rho_{\rm atm})$ as \cite{khoury}.
Then, the asymptotic scalar field profile is obtained by linearizing the KG equation up to first order around spatial infinity,
\bea
\phi''+\frac{2}{r} \phi'= \mathcal{M}^2 \left(\phi-\phi_{\infty}\right),
\eea
with $\mathcal{M}^2=\left.\dd^2 V_{\rm eff}/\dd\phi^2\right|_{\phi=\phi_{\infty}}$, which admits the Yukawa profile 
solution $(\mathcal{M}^2>0)$
\be
\phi=\phi_{\infty}+\frac{\mathcal{C}{\rm e}^{-\mathcal{M}r}}{r},
\label{asympt}
\ee
with $\mathcal{C}$ the constant of integration. Since the KG equation is of second order and the parameter 
$\mathcal{C}$ is to be determined, three boundary conditions are needed.
They are provided by the regularity condition on the scalar field derivative at the origin $\phi'(r=0)=0$ and by the 
asymptotic behavior of $\phi$ and $\phi'$ given by Eq.\eqref{asympt} at the end of the 
integration interval. For the multipoint bvp method, the continuity of $\phi$ and $\phi'$ are imposed at 
the interfaces of each region (6 conditions) while the profile is guaranteed to be continuous for arctan profiles of density.
The density and size of each region are reported in Table~\ref{tab:expconf}.
The two numerical methods have been checked to be in agreement with each other.
Their applicability to the various regimes and their limitations in the deep thin-shell regime 
will be discussed in Sec.~\ref{secChamelNum}. We already point out that this numerical method enables to properly account 
for the effect of neighboring matter on the chameleon fields and can be easily generalized to other experiments, possibly more sensitive (in the limit of spherical symmetry).


\section{Analytical approach}
\label{sec5}

In this section we reproduce the main steps of \cite{Burrage} and derive analytically the chameleon field profile 
in the spherically symmetric and static regime for a two-region model (the source mass and the vacuum chamber). 
In the next section, the validity of the various assumptions will be analyzed and the analytical approximations will be compared to the exact numerical results, for the
two chameleon potentials of Table~\ref{tab:models}. 
For the sake of simplicity, we assume in this section that $\alpha=1$. 

Assuming $A(\phi)=1$, the field value at the minimum of the effective potential written in Eq.~(\ref{eq:KG}), and the field mass around it, 
are respectively given by
\bea
\phi_{\rr{min}} = \left( \frac{\Lambda^5 M}{\rho} \right)^{1/2}, \hspace{0.5cm} m_{\rr{min}} = \sqrt 2 \left( 
\frac{\rho^3}{\Lambda^5 M^3} \right)^{1/4}.
\eea
The case where the effect of $A(\phi)$ is important will be discussed in Sec.~\ref{secChamelNum}.
For a two-region model the density $\rho$ is either the source mass density $\rho_{\rr A}$ or the density in 
the vacuum chamber $\rho_{\rr{v}}$.  

Four different regimes can be identified, depending on whether the field reaches the effective potential minimum or not: 
(1) the field does not reach the minimum of the effective potential in any region, (2) the field reaches the minimum 
in the vacuum chamber but not in the source mass, (3) the field reaches the minimum in the 
source mass but not in the vacuum chamber,  (4) the field reaches the minimum both inside the test 
mass and the vacuum chamber.   Cases (1) and (2) were 
referred as the \textit{weakly perturbing regime} in Ref.~\cite{Burrage}, whereas (3) and (4) were referred as 
\textit{strongly 
perturbing}. Below we consider those four cases separately, as in Ref.~\cite{khoury}.   In principle, one should also distinguish between the cases where the field reaches $\phi_{\rr{min}} $ inside the chamber wall, or not.  When lowering $M$, depending on the central mass density and size, on the chamber wall density and thickness, $\phi_{\rr{min}} $ can be reached first inside the central mass or inside the chamber walls.  Nevertheless, for the considered experimental set-up, the wall and the central mass have similar densities and sizes, and so those two cases will not be distinguished in the following. 

\subsubsection{$\phi(r=0) \neq  \phi_{\rr{min}}(\rho_A) $ and $\phi(R_A <r <L) \neq  \phi_{\rr{min}}(\rho_{\rr 
v}) $}

Within the test mass the field does not reach the attractor that is the minimum of the effective potential.  Since $\rho_{\rr v} < 
\rho_{\rr{atm}} 
< \rhoA$,  the second term in the effective 
potential dominates, $V_{\rr{eff}} \simeq \phi \rhoA /M$.   The KG equation can be solved inside the mass imposing that the 
field profile is regular at the origin, which gives
\bea \label{eq:KGinA}
\phi = D + \frac{\mA r^2}{8 \pi M \RA^3}~,
\eea
where $D$ is an integration constant that can be fixed by matching $\phi$ and $\phi'$ to the field solution in the 
vacuum chamber at $ r=R_A $.  Inside the vacuum chamber the 
field does not reach the attractor value.   Let us denote $\phibg$ the value that would take the field at the center of 
the chamber in the absence of the source.  Then one 
can consider an harmonic expansion of the potential 
\bea \label{eq:Veff}
\Veff(\phi) \simeq \Veff(\phibg) + \frac {\mbg^2}{2} (\phi - \phibg)^2~,
\eea
higher order terms being subdominant.  One can solve the KG equation assuming that the field profile 
decays at infinity.  This gives
\bea \label{eq:KGinvacuum}
\phi(r) = \phibg + \frac{\alpha}{r} \rr e^{- \mbg r}~.
\eea
Note that at $r=R_A$, one has $\mbg R_A \ll 1$ for typical experimental parameters and thus $\phi(R_A )\simeq \phibg + 
\alpha/R_A$.  After matching, one finds the field 
profile in case (1)
\bea \label{eq:case1}
\phi^{(1)}(r) = \phibg &-& \frac{\mA}{8 \pi \RA M} \times \left[ \left( 3 - \frac{r^2}{\RA^2}  \right) \Theta(\RA - 
r)\right. \non\\
&+&\left. \left( 2 \frac{\RA}{r} \rr e^{-\mbg r} \right) \Theta(r- \RA) \right],
\eea
where $\Theta$ is the Heaviside function.  Therefore the effect of the mass is to deepen the field profile, by a quantity $3 \mA /(8 \pi \RA M) \ll \phibg$ 
at $r=0$.  By definition, case (1) is valid as 
long as $|\phibg - \phi^{(\rr 1)}(r=0)| \ll \phibg$.
Outside the mass, the difference $|\phibg - \phi|$ decreases like $\propto 1/r$ for realistic experimental 
configurations where the exponential decay factor can be 
neglected. 

A subtlety arises in the evaluation of $\phibg$, which in Ref.~\cite{Burrage} was either the attractor in the vacuum, either related to the chamber 
size\footnote{$\rho_{\rr v}$ is 
much lower than the wall density $\rho_{\rr w}$ where the field was assumed to reach its attractor 
$\phi_{\rr{min}}\left(\rho_\rr{w}\right)$.  Thus the 
first term of $\Veff$ in Eq.~(\ref{eq:Veff}) dominates  
the KG equation inside the chamber, which can be solved to get $\phibg $ as a function of 
the size of the vacuum chamber.  However, behind 
this calculation is hidden the assumption that the field reaches $\phi_{\rr{min}}\left(\rho_\rr{w}\right)$ in the 
wall, which is not valid in case (1) in most of the parameter space.},
under the assumption that the scalar field reaches the minimum of the effective potential inside the vacuum chamber wall.
This assumption is actually not valid in case (1) 
because $\rhoW\sim\rhoA$, and because the wall 
thickness is about the radius of the test mass.  
So in most of the parameter space corresponding to case (1), the scalar field does not reach its attractor inside the wall.
As a result, $\phibg$ is better approximated by $\phi_\rr{min}(\rhoATM)$.
Numerical results will highlight the effects of the chamber wall on the scalar field profile.
Even if the background field value 
has no effect on the acceleration itself, this result is important because it 
changes the region in the parameter space in which case (1) applies: it is extended to lower values 
of $M$, as developed thereafter. 

\begin{figure}
\begin{center}
\includegraphics[scale=0.45, trim= 230 0 240 0, clip=true]{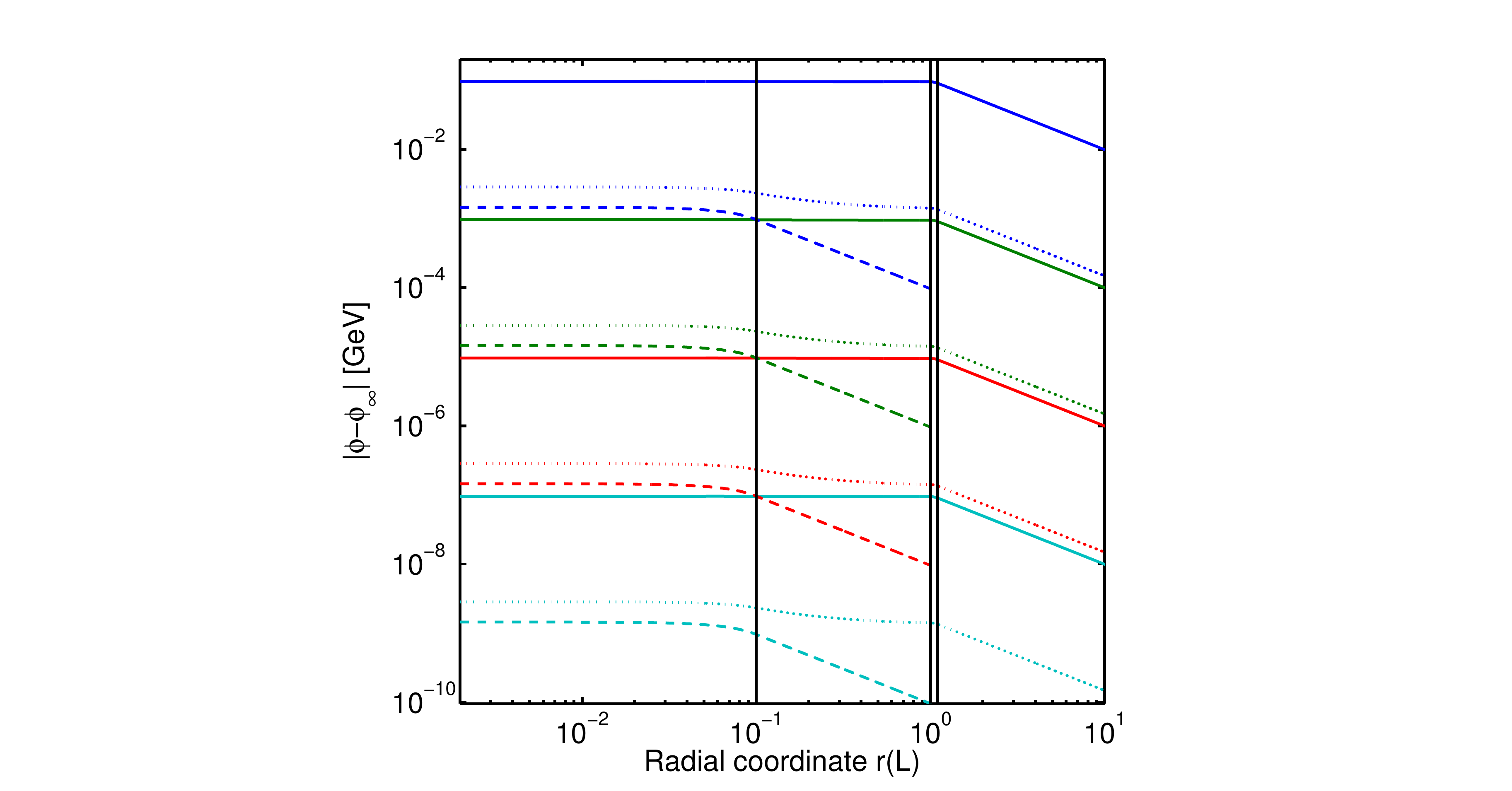}
\caption{Numerical and analytical scalar field profiles (respectively solid and dashed lines) in absolute value for 
various $M$ listed in Table~\ref{tab:profiles}. The numerical profiles obtained when lowering the wall density to   $\rhoW=5\times10^{-19}<\rhoA$,
are drawn in dotted line, in order to illustrate that the wall density can perturb more or less importantly the field profile.   
Vertical lines mark out the four regions: the test mass, the chamber, the wall and outside the chamber.}
\centering
\label{plot_profile_field_k}
\end{center}

\begin{center}
\includegraphics[scale=0.45, trim= 230 0 240 0, clip=true]{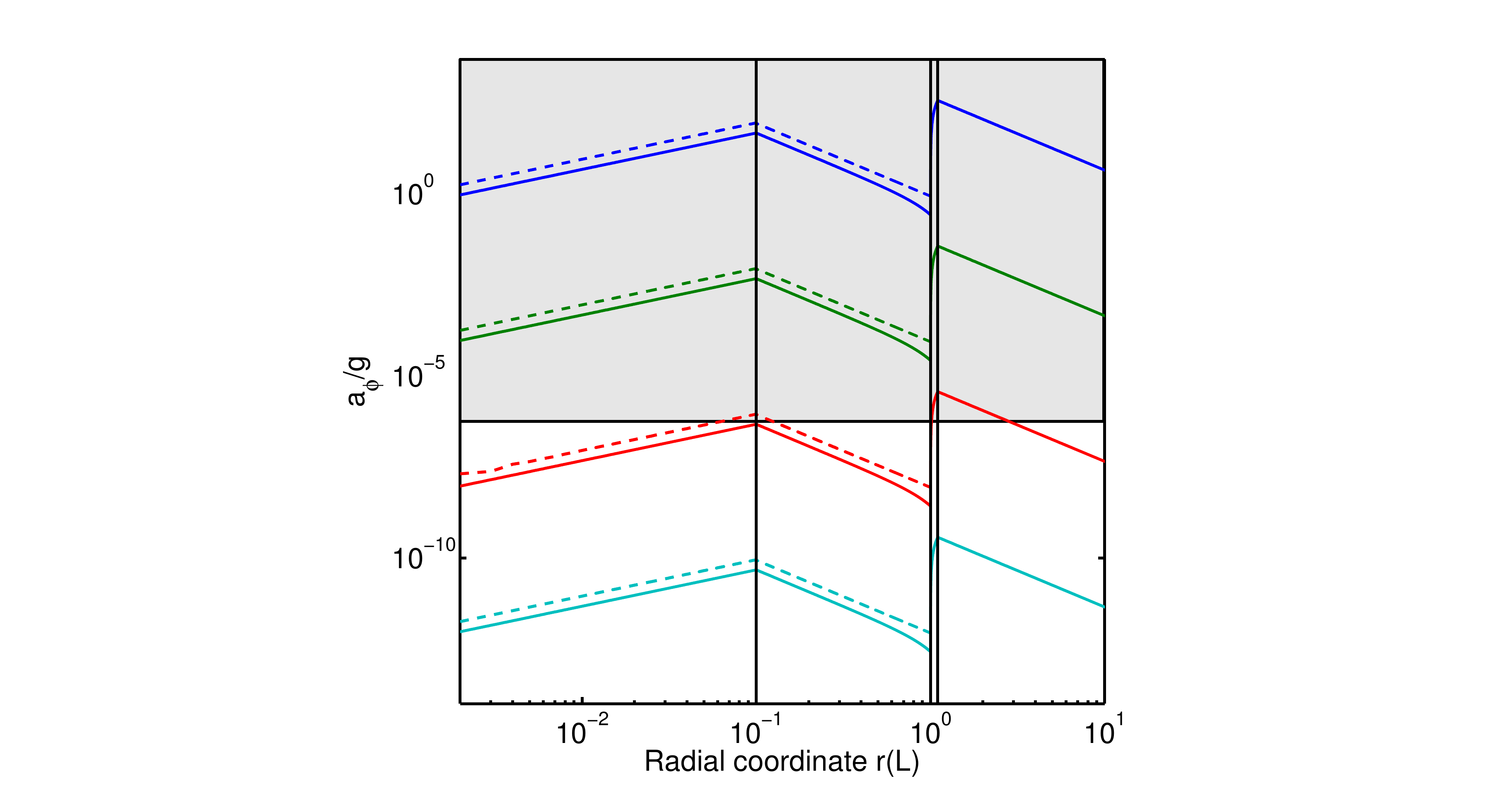}
\caption{Numerical and analytical profiles (respectively solid and dashed lines) of the acceleration $a_\rr\phi/g$ with $g$ the Earth gravitational acceleration, for $M$ values listed in Table~\ref{tab:profiles}. 
Vertical lines mark out the four regions (test mass, chamber, wall and exterior).}
\centering
\label{plot_profile_acc_k}
\end{center}
\end{figure}

The analytical field profile and the induced acceleration $a_\phi = \partial_r \phi / M$ have 
been plotted on Figs.~\ref{plot_profile_field_k} and \ref{plot_profile_acc_k} respectively for various
values of $M$ reported in Table~\ref{tab:profiles}. 

\begin{table} 
\begin{tabular}{|c|c|c|c|} 
\hline
Color & M [GeV] & $a_\phi/g$ (\textit{near}) & $a_\phi/g$ (\textit{far}) \\ 
\hline
\multicolumn{4}{|c|}{Chameleon-1, \textit{weakly perturbing}: Figs.~\ref{plot_profile_field_k},~\ref{plot_profile_acc_k}  }  \\
\hline
Blue & $10^{13}$&  $1.3 \times 10^{1}$ & $2.8\times 10^0$ \\ 
Green & $10^{15}$  &  $1.3\times 10^{-3}$ & $2.8\times 10^{-4}$\\ 
Red & $10^{17}$ &  $1.3\times 10^{-7}$ & $2.8\times 10^{-8}$\\ 
Light blue & $10^{19}$ &  $1.3\times 10^{-11}$ & $2.8\times 10^{-12}$\\ 
\hline
\multicolumn{4}{|c|}{Chameleon-1, \textit{thin-shell}: Figs.~\ref{plot:profile:field_thinshell1},~\ref{plot:profile_acc_thinshell1}  }  \\  \hline
Blue & $10^{8}$ &  $5.8 \times 10^{9}$ & $1.4 \times 10^{8}$ \\ 
Green & $10^{9}$ &  $5.2 \times 10^{8}$ & $5.7 \times 10^{6}$   \\
Red & $10^{10}$ &    $1.9 \times 10^{7}$ & $-4.4 \times 10^{6}$  \\ 
Light blue & $10^{11}$ &  $2.5 \times 10^{5}$  & $5.5 \times 10^{4}$  \\ 
\hline
\multicolumn{4}{|c|}{Chameleon-2, \textit{thin-shell}: Figs.~\ref{plot:cham2_field},~\ref{plot:cham2_acc}  }  \\ \hline
Blue & $10^{14}$&  $ 5.2 \times 10^{-7}$  & $ 1.5 \times 10^{-8}$ \\ 
Green & $10^{15}$ & $ 5.2 \times 10^{-8}$ & $ 1.5 \times 10^{-9}$ \\ 
Red &$10^{16}$ & $ 5.2 \times 10^{-9}$ & $ 1.5 \times 10^{-10}$  \\ 
Light blue & $10^{17}$ &  $ 5.2 \times 10^{-10}$ &$ 1.5 \times 10^{-11}$  \\ 
Purple &$10^{18}$ & $ 5.3 \times 10^{-11}$ &$ 2.4 \times 10^{-12}$  \\ 
Beige&$10^{19}$ & $ 4.6 \times 10^{-12}$ &$ 6.8 \times 10^{-14}$  \\
\hline
\end{tabular}
\caption{Properties of the numerical scalar field and acceleration profiles
for the two models in the different regimes.}
\label{tab:profiles}
\end{table}


The acceleration induced by the scalar field gradient inside the vacuum chamber is well approximated by
\be
 a_\phi \approx \frac{\mA}{4 \pi  M^2 r }  \left( \frac{1}{r} + \mbg  \right). 
\ee
Since $\mbg r \ll 1$ for realistic laboratory experiments, the acceleration is independent of $\Lambda$ and thus one can constrain directly the 
value of $M$. This is the reason why, as we will show in the following, the 
power-law of the potential has no effect on the acceleration as long as $|A(\phi)-1|\ll1$.

\subsubsection{$\phi(0) \neq  \phi_{\rr{min}}(\rho_A) $ and $\phibg = \phi_{\rr{min}}(\rho_{\rr v}) $}

When the size of vacuum chamber is larger than the characteristic distance over which the field reaches the minimum of 
the potential, that is when
\bea \label{eq:case2condition}
L \gg \frac{1}{m_{\rr{min}}(\rho_{\rr v})} = \left(\frac{\Lambda^5 M^3}{4 \rho_{\rr v}^3}\right)^{1/4}~,
\eea
 the field profile is still governed by Eq.~(\ref{eq:case1}).  However the value of $\phibg$ is now simply 
$\phi_{\rr{min}} (\rho_{\rr v}) $.   In the case of the bare 
chameleon potential $V(\phi) = \Lambda^5 / \phi$, one has $ \Lambda \simeq 2.6 \times 10^{-6} \GeV $ in order to 
reproduce the late-time accelerated expansion of the 
Universe.  For typical vacuum densities and chamber sizes, e.g. those reported in Table~\ref{tab:expconf}, one 
finds that this 
regime would occur when $M\lesssim 10^{-6} \GeV$.  
This does not correspond anymore to the weakly perturbing regime requiring $\phibg \gtrsim m_A/ (4\pi R_A M)$, which 
gives $M 
\gtrsim 2 \times 10^9 \GeV$ in our fiducial experimental setup.  In the case of the potential  $V(\phi) = \Lambda^4 
(1+ \Lambda / \phi)$, $\Lambda \simeq 10^{-12}\GeV $ 
is the cosmological constant.   It results that the field 
in the chamber is expected to reach $\phi_{\rr{min}}(\rho_{\rr v})$ only if $M\lesssim 10^5 \GeV$.   There again this is far from 
the regime where the test mass perturbs only weakly the field, valid when $M \gtrsim 10^{20} \GeV$, i.e. in the 
super-Planckian regime.

 \subsubsection{$\phi(0) = \phi_{\rr{min}}(\rho_A) $ and  $\phi(R_A <r <L) \neq  \phi_{\rr{min}}(\rho_{\rr 
v}) $}
 
 In case (3) the field reaches $\phi_{\rr A} \equiv \phi_{\rr{min}}\left(\rho_\rr{A}\right)$ inside 
the test mass.  One can 
define a radius $S$ such that $\phi(S) = \phi_{\rr A} 
(1+ \epsilon) $ with $0<\epsilon \ll1 $. For $S < r < \RA$, the density term dominates in 
$V_{\rr {eff}}$ and the solution of the linearized KG equation is given by  
 \bea
 \phi = D + \frac{C}{r} + \frac{\mA r^2}{8 \pi M \RA^3}~,
 \eea
 which is the same as Eq.~(\ref{eq:KGinA}) but with a non-vanishing integration constant $C$.  Outside the test 
mass, the field still obeys to 
Eq.~(\ref{eq:KGinvacuum}).  After matching $\phi$ and $\phi'$ at $r = \RA$ and $\phi$ at $r=S$,  the 
integration constants $\alpha$, $D$ and $C$ can be fixed. The resulting field profile in case (3) reads~\cite{Burrage}

\begin{widetext}
 \bea \label{eq:case3}
 \phi^{(3)}(r) = \left\{
 \begin{split}
 & \phi_{\rr A}~, & 
 &  r < S, & \\
 &  \phi_{\rr A}+  \frac{\mA}{8 \pi \RA^3 M r} \left( r^3 - 3 S^2 r + 2 S^3 \right) ~, &   
 &  S<r<\RA, &  \\
 &   \phibg - \frac{\mA }{4 \pi M r} \rr e^{-m_{\rr{bg}} r} \left( 1 - \frac{S^3}{\RA^3}  \right)~, &
 &  r>\RA, &     
 \end{split}
 \right.
 \eea
\end{widetext}

 with the radius 
 \bea  \label{eq:radius_tt}
 S \equiv \RA \sqrt{1 - \frac{8 \pi M \RA \phibg}{3 \mA}}
 \eea

 being such that one has typically $(\RA - S) /\RA  \ll 1$, corresponding to the thin shell regime.   
 The induced acceleration is well approximated ($\mbg\RA\ll1$) by
 \be
  a_\phi \approx \frac{\mA}{4 \pi  M^2 r^2 }  \left( 1 - \frac{S^3}{\RA^3}  \right) \simeq \frac{ \RA \phibg}{M r^2 } 
 \ee
 and contrary to case (1), it is related to the value of $\phibg$. If the wall 
is sufficiently large, then the field reaches $\phi_\rr{min}(\rhoW)$ and so the calculation of $\phibg$ in 
Ref.~\cite{Burrage} is valid, giving
\be  \label{phibgSP}
\phibg \simeq 0.69 \left(\Lambda^5 L^2\right)^{1/3}
\ee
for a spherical chamber. Compared to the case (1), the induced acceleration therefore does not 
depend only on $M$ but also on $\Lambda$ and on the size of the vacuum chamber $L$. 
When $\Lambda$ is set to the cosmological constant and $L$ to the fiducial value reported in Table~\ref{tab:expconf}, one finds that the experiment of~\cite{khoury} constrains the coupling parameter down to $M \sim 10^{15} \GeV$. The above calculation does not involve the power-law index $\alpha$ (apart indirectly via $m_{\rr{bg} }$, but there is no effect in the limit $m_{\rr{bg} } r \ll1$).  Therefore it is expected that the predictions are independent of $\alpha$, as long as $|A(\phi) - 1 | \ll 1$.  

\begin{figure}
\begin{center}
\includegraphics[height=85mm, trim= 240 0 250 0, clip=true]{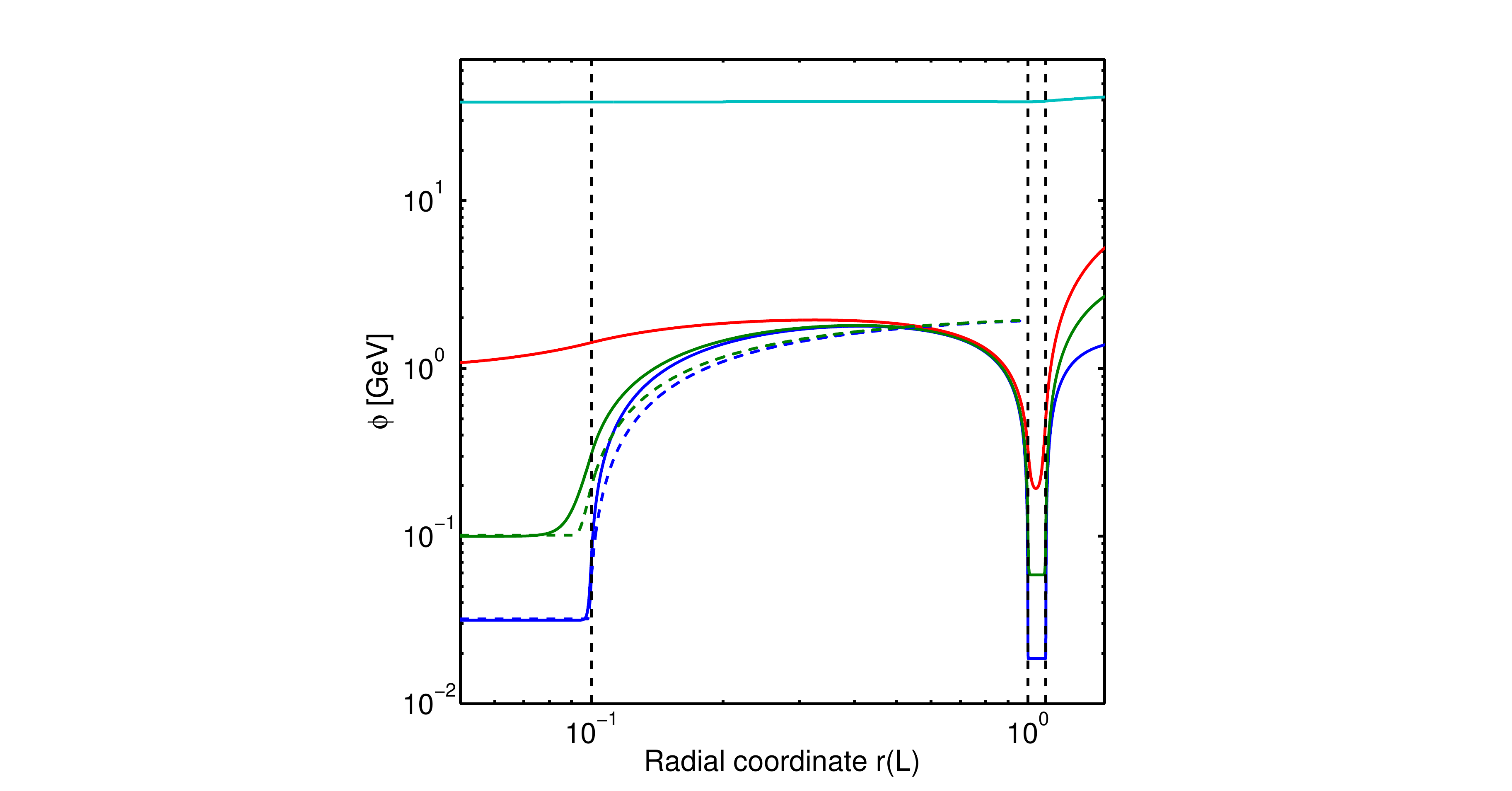}
\caption{Numerical results (solid lines) and analytical approximation (dashed lines) for the scalar field profile of the Chameleon-1 model, in the \textit{strongly perturbing} (thin shell) regime, for $\Lambda = 2.6 \times 10^{-6} \GeV$ and values of the coupling $M$ listed in Table~\ref{tab:profiles}.  Differences between the two-region and four-region models are non-negligible inside the chamber, especially at the vicinity of the wall.
Vertical lines mark out the four regions (test mass, chamber, wall and exterior).}
\centering
\label{plot:profile:field_thinshell1}
\end{center}

\begin{center}
\includegraphics[height=85mm, trim= 240 0 250 0, clip=true]{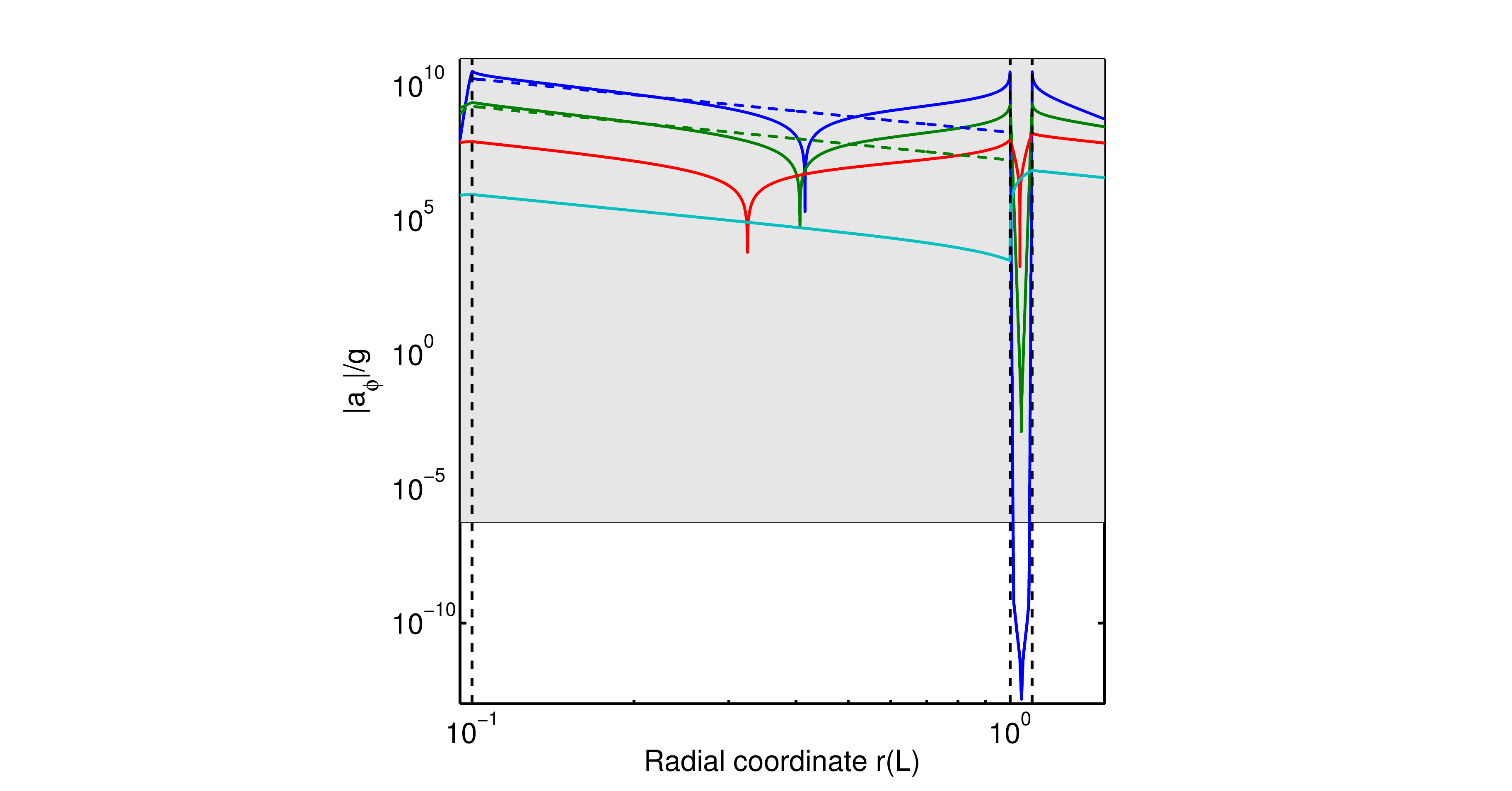}
\caption{Numerical results (solid lines) and analytical approximation (dashed lines) for the acceleration $|a_\rr\phi| / g$ profile, for the same model and parameters as in Fig.~\ref{plot:profile:field_thinshell1}. The numerical profile for the four-region model shows that from the middle of the chamber to the wall, the acceleration becomes negative and increases in magnitude. Vertical lines mark out the four regions (test mass, chamber, wall and exterior).}
\centering
\label{plot:profile_acc_thinshell1}
\end{center}
\end{figure}

Analytical field profiles and induced    
accelerations for case (3) are represented on Figs.~\ref{plot:profile:field_thinshell1} and \ref{plot:profile_acc_thinshell1} for the bare potential $V(\phi) = \Lambda^5 / \phi$ (Chameleon-1), and on Figs.~\ref{plot:cham2_field} and \ref{plot:cham2_acc} for the potential $V(\phi) = \Lambda^4 (1+ \Lambda / \phi ) $,  
for several values of $M$ reported in Table~\ref{tab:profiles}.   Those are found to be in good agreement with the 
numerical results, except close to the wall where important deviations are found.

In the strongly perturbig regime, the reliability of the theory is questionable. Indeed the quantum corrections, either in the matter and the chameleon sector must remain small. Most of the parameter space reachable by the experiment proposed in \cite{khoury} belongs to this regime. Following \cite{Upadhye:2012vh} the underlying instabilities are harmless and the classical analysis is trustable, keeping in mind that quantum corrections can become large at very small scales.
However since the aim of this paper consists of modelling how the environement can affect the analytical results derived for the classical field, we also provide numerical forecasts in the questionable strongly perturbing regime. 
Nevertheless we did not explore the deeply strong regime but focus on the transition between the two regimes, where the numerical computations allow to follow the smooth evolution of the field and acceleration profiles whereas analytical assumptions break.  Our computations show that the analytical estimations are recovered once in the strong regime, and that they are quite reliable, at least classically.  The underlying quantum aspects are beyond the scope of this paper.  


\subsubsection{$\phi(0) =  \phi_{\rr{min}}(\rho_A) $ and $\phibg = \phi_{\rr{min}}(\rho_{\rr v}) $}

In case (4) the field profile is governed by Eq.~(\ref{eq:case3}) since the field reaches the effective potential 
minimum at the center of the test mass. However, as long as the 
condition Eq.~(\ref{eq:case2condition}) is satisfied, $\phibg = \phi_{\rr{min}}\left(\rho_{\rr v}\right)$.  For the 
bare potential $ 
V(\phi) = \Lambda^5 /\phi $, case (4) takes place when $M \lesssim 10^{-3 } \GeV$, whereas for the potential $ V(\phi) 
= \Lambda^4 (1+ \Lambda /\phi) $ one needs $M 
\lesssim 10 \GeV$ in order to be in the strongly perturbing regime inside the test mass.   Therefore the case (4)  is irrelevant 
for values of $\Lambda$ motivated by cosmology and realistic experimental configurations. 

\section{Four-region model: numerical results}\label{secChamelNum}

\subsection{Weakly perturbing regime}
We discuss the weak field regime for Chameleon-1 exclusively since only the case (3) of the previous 
section (strongly perturbing regime) is relevant for values of $M$ below the Planck scale in the Chameleon-2 model.   
Allowing superplanckian values, one actually would recover the regime where the field is only weakly perturbed by the 
central mass, but the induced acceleration would be far too low for being observable with future experiments.   

On Fig.~\ref{plot_profile_field_k}, $|\phi-\phi_\infty|$ is 
represented for various $M$ values of the Chameleon-1 model, $\alpha=1$ and $\Lambda=2.6\times 10^{-6}$~GeV being 
fixed, which corresponds to case (1) discussed in the previous section. Inside the test mass the scalar field is 
constant 
but the numerical profile roughly differs 
by two orders of magnitude compared to the analytical approximation. This difference is induced by the effect of the wall, which 
enhances $|\phi-\phi_\infty|$ in the vacuum chamber: the wall tends to stabilize the scalar field and gives it a kick 
right outside the wall shell.  Note that the importance of the effect depends on the wall density and thickness.  By setting $\rho_{\rr w} \ll \rho_{\rr A}$ (or by reducing the wall thickness), one tends to recover the analytical profile. Outside the chamber, the scalar field follows the Yukawa profile as imposed by the 
asymptotic behavior in Eq.~\eqref{asympt}. 

The acceleration $a_\phi/g=\partial_r\phi/(M g)$ with $g$ the Earth gravitational acceleration, is plotted 
on Fig.~\ref{plot_profile_acc_k}. The general behavior of the acceleration profile obtained numerically 
does not differ significantly from the 
analytical approximation.  However the chamber wall affects the amplitude of the profile 
with a difference growing up to one order of magnitude at the vicinity of 
the wall. This result illustrates how important it is to take into account the four regions modeling the experiment, 
in the weakly perturbing regime, in the view 
of establishing accurate constraints from atom-interferometry experiments.   
Given the experimental constraint on the acceleration 
$a_\rr{exp}/g<5.6\times 10^{-7}$ \cite{khoury}, we find that the Chameleon-1 model is excluded at 95\% C.L. for $\alpha=1$ and 
$M < 7 \times 10^{16}$ GeV (see Table~\ref{tab:profiles}).    An experiment controlling systematics to probe 
$a_\phi/g\lesssim 10^{-12}$ would rule out the model up to the Planck scale.   This is close to the value $a_\phi/g \sim 
10^{-11}$ given in~\cite{Burrage} as a reachable sensitivity.

\begin{figure}
\begin{center}
\includegraphics[scale=0.45, trim= 220 0 240 0, clip=true]{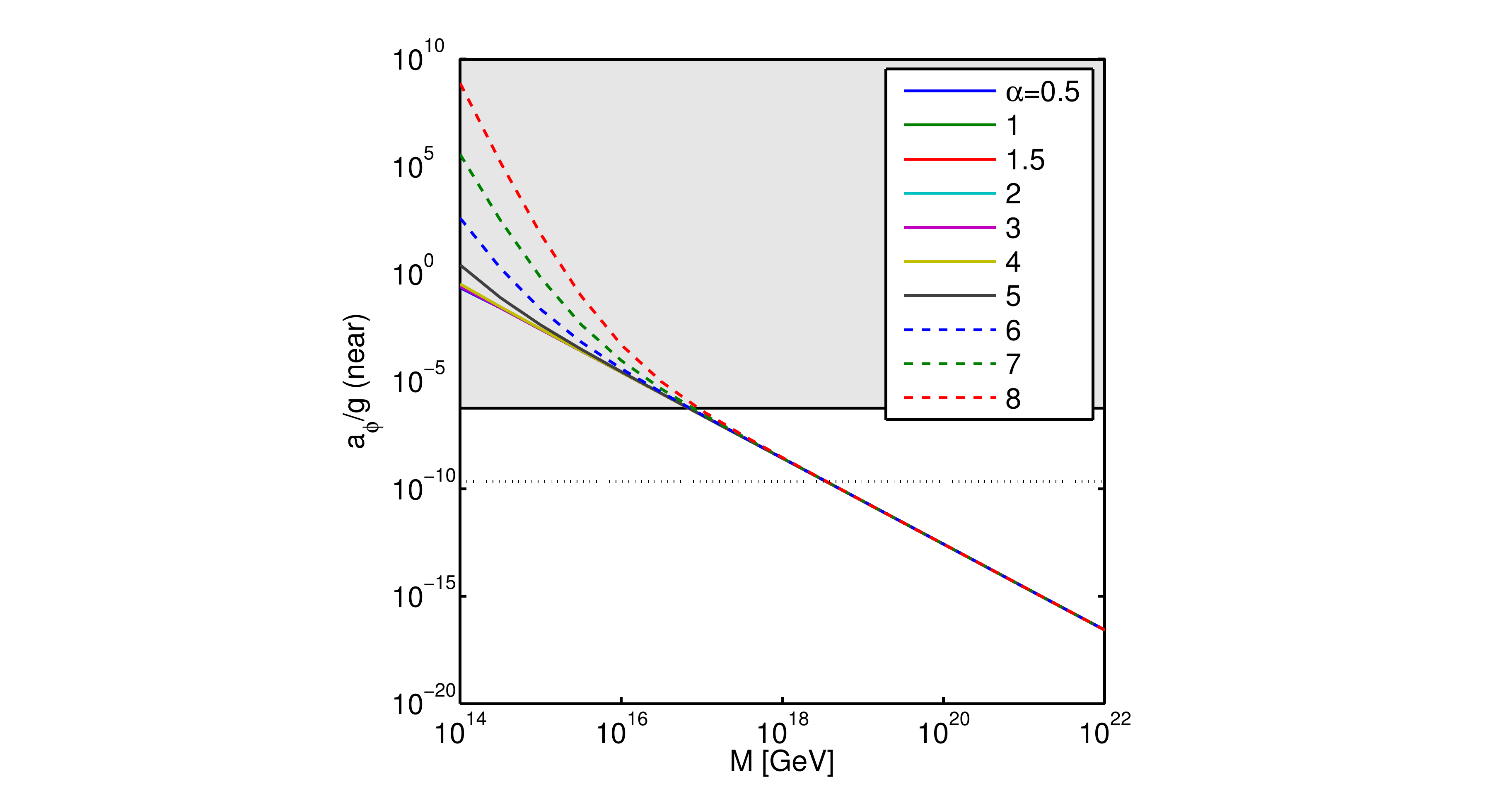}
\caption{Forecast for the normalized acceleration $a_\phi/g$ measured in the "near" position, i.e. 
$8.8$ mm far from the test mass, for various $M$ and $\alpha$. Dotted line represents the gravitational acceleration due to the test mass.}
\centering
\label{simu_near}
\end{center}
\end{figure}

On Fig.~\ref{simu_near}, the parameter space of $\left(\alpha,M\right)$ is explored for the acceleration 
measured in the \textit{near} position (see Sec.~\ref{sec3}). Deviations between \textit{near} and \textit{far} positions are 
tiny (see Table~\ref{tab:profiles} for an order of magnitude).
A universal behavior with respect to $\alpha$ is observed for $M>10^{17}$~GeV. Given the current bounds 
on $a_\phi / g$, we thus find that the coupling parameter $M$ is constrained identically for Chameleon-1 models, independently of the $\alpha$ power-law parameter.  Future experiments will not be able to distinguish between the various power-law potentials.  
The universal behavior is broken at lower values of $M$ because the assumption 
$|A(\phi)-1|\ll1$ is not valid and the analytical approach cannot be trusted anymore.

\subsection{Strongly perturbing regime}\label{sec:strong}

Probing the deep thin-shell regime, i.e. when $(R_{\rr {A,w}} - S_{\rr {A,w}})\ll R_{\rr {A,w}} $ (thin-shell radius of the test mass or of the chamber wall), 
 is very challenging numerically.  Up to some point, it is nevertheless possible to track the solution and to check the validity of the analytical estimations, typically using mesh refinement methods.  The numerical treatment also allows to probe the smooth transition between the weakly and strongly perturbed cases.  

\begin{table} 
\begin{tabular}{|c|c|c|} 
\hline
Color & $\rhoA~\left[\rr{GeV}^4\right]$ & $\rhoW~\left[\rr{GeV}^4\right]$  \\
\hline
Blue & $1.0\times 10^{-20}$& $1.0\times10^{-20}$  \\
Green & $2.5\times 10^{-20}$  & $2.5\times 10^{-20}$\\
Light blue & $5.0\times 10^{-20}$ & $5.0\times 10^{-20}$ \\
Purple & $7.5\times 10^{-20}$ & $7.5\times 10^{-20}$ \\
Beige  & $5.0\times 10^{-19}$ & $7.5\times 10^{-20}$ \\
Red & $1.2\times 10^{-17}$ & $7.5\times 10^{-20}$ \\
\hline
\end{tabular}
\caption{Densities inside the test mass $\rhoA$ and the wall $\rhoW$ for the numerical scalar field and acceleration 
profiles of Figs.~\ref{plot:density_field} and \ref{plot:density_acc}.}
\label{tab:density}
\end{table}

\begin{figure}
  \includegraphics[scale=0.42, trim=220 0 240 0, clip=true]{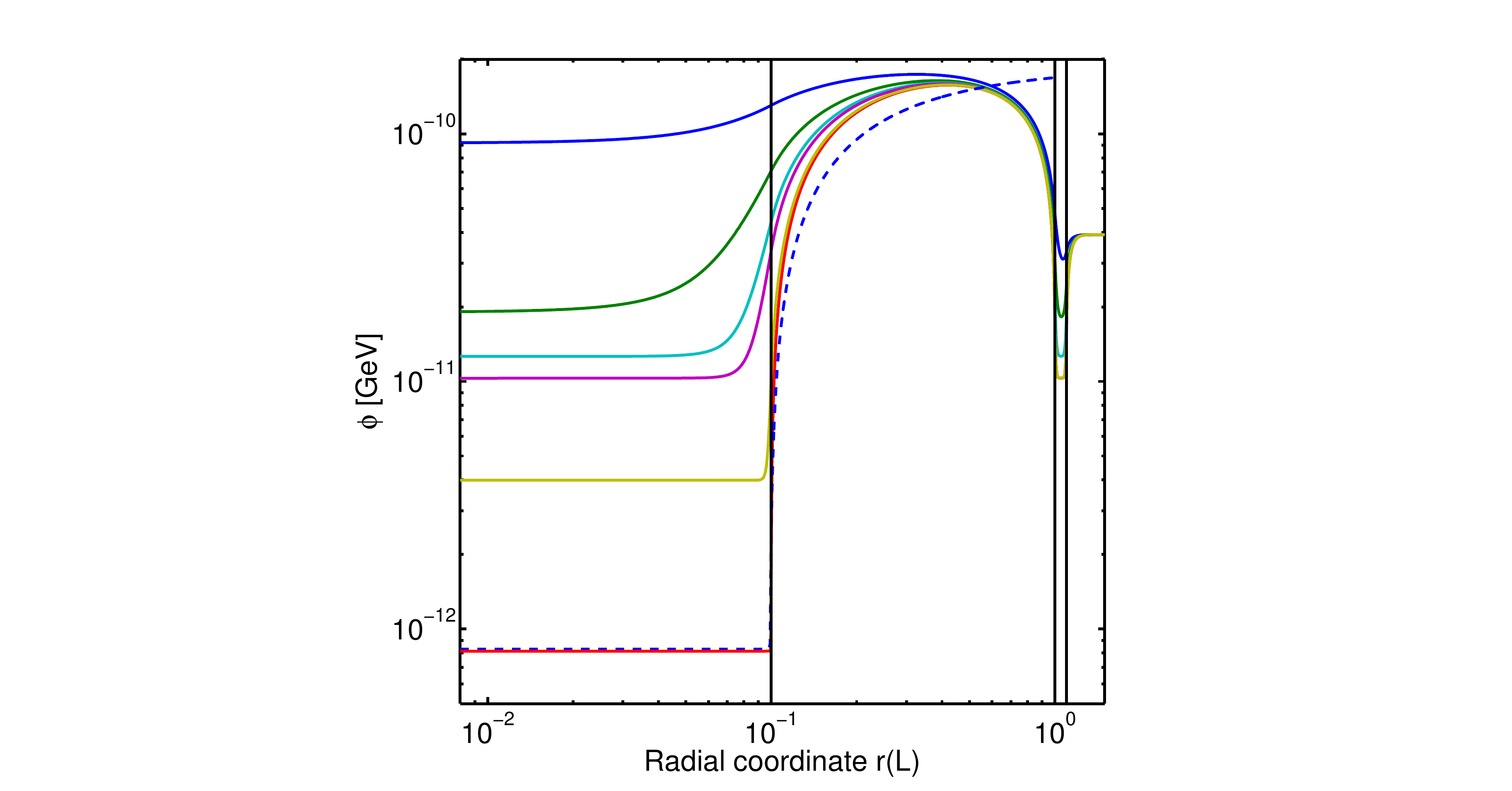}
  \caption{Numerical results (solid lines) and analytical approximation (dashed line) for the scalar field profile of 
  the Chameleon-2 model, for various $\rhoA$ and $\rhoW$ reported in Tab.~\ref{tab:density}, $M=10^{17}$ GeV being 
  fixed.}
  \label{plot:density_field}
%
  \includegraphics[scale=0.42, trim=220 0 240 0, clip=true]{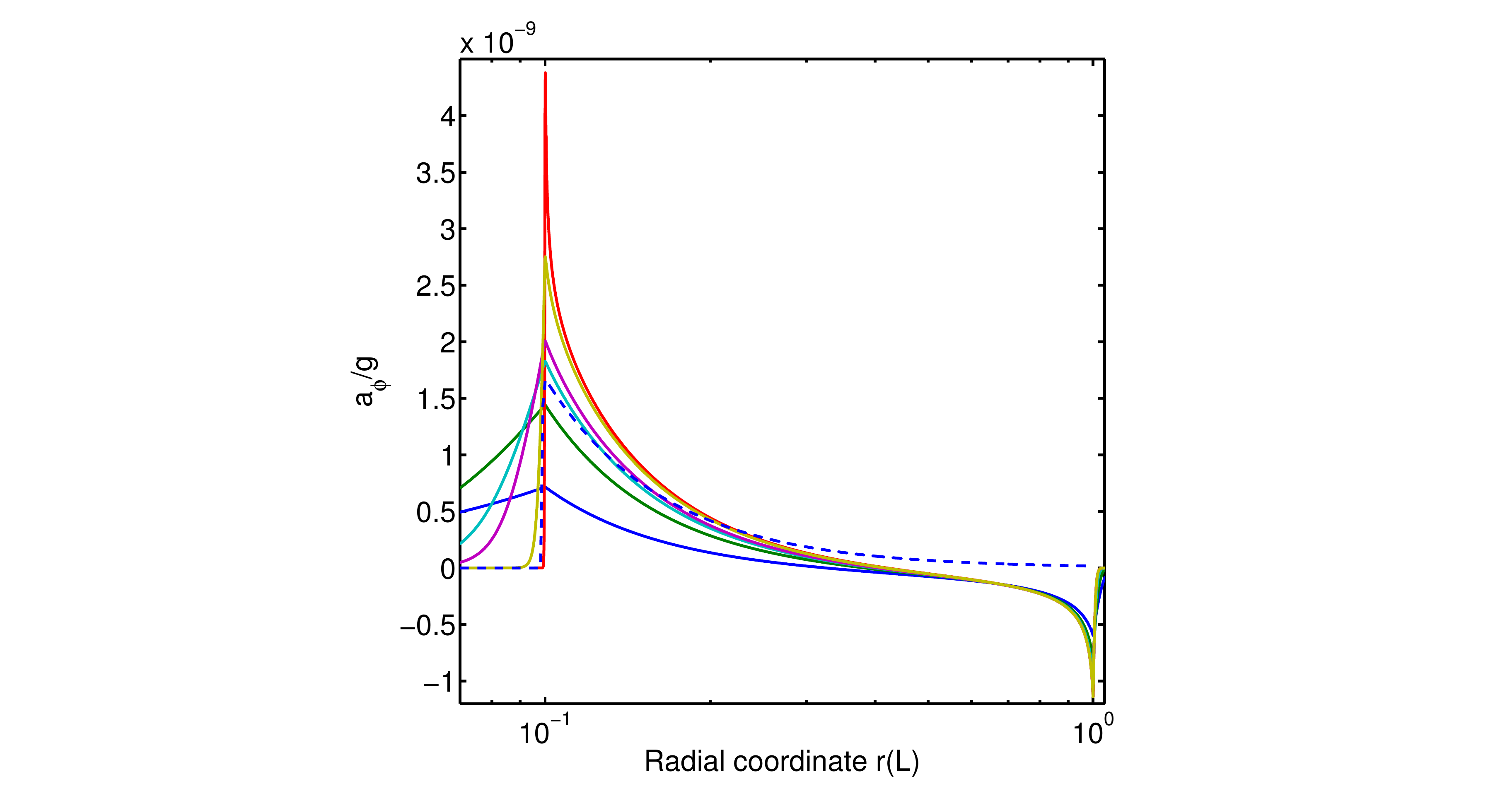}
  \caption{Numerical results (solid lines) and analytical approximation (dashed line) for the acceleration $a_\phi/g$ 
  profile of the Chameleon-2 model, for various $\rhoA$ and $\rhoW$ reported in Tab.~\ref{tab:density}, $M=10^{17}$ GeV 
  being fixed.}
  \label{plot:density_acc}
\end{figure}

\subsubsection{Chameleon-1}
Even if one has already predicted analytically that the acceleration would be excluded, the field and acceleration profiles 
have been also computed for parameters corresponding to the strongly perturbed regime, referred as Case (3) in the 
Sec.~\ref{sec5}.  Those are represented on Figs.~\ref{plot:profile:field_thinshell1} 
and~\ref{plot:profile_acc_thinshell1} and compared to analytical predictions for several values of $M$.  Our numerical 
method also allows to probe the transitory regime

As expected given that $\rho_{\rr A} \lesssim \rho_{\rr w}$ with similar test mass radius and wall thickness, when lowering M, the field reaches first the potential minimum $\phi_{\rr{min}}(\rho_{\rr w})$ inside the wall, and then $\phi_{\rr{min}}(\rho_{\rr A})$ within the test mass, over a very thin radius.  Inside the vacuum chamber, one observes that the field roughly reaches
the amplitude of $\phibg$ given by Eq.~(\ref{phibgSP}), which validates the calculation of~\cite{Burrage}.  
In the vicinity of the chamber wall, however, the acceleration changes its sign and becomes negative, 
with a comparable magnitude with the acceleration close to the source mass.   
This effect could be helpful experimentally to discriminate between a signal of modified gravity and systematic errors, by performing measurements of the acceleration at several key positions of the chamber.


\subsubsection{Chameleon-2}  \label{sec:strong_field_cham_2}

For the Chameleon-2 model and the considered experimental set-up, 
it has been impossible to track numerically the thin-shell regime up to the point where the acceleration would have been large enough to be observed in laboratory experiments.  
Nevertheless, the field and acceleration profiles are represented on Figs.~\ref{plot:density_field} and~\ref{plot:density_acc}, for $M=10^{17} \rr{GeV}$ and increasing values of $\rho_{\rr w}$ and $\rho_{\rr A}$.   The attractor field values within the test mass and the wall are reached progressively and the field variations at the borders between the four regions become more steep, as expected given that 
$(R_{\rr {A,w}} - S_{\rr {A,w}}) / R_{\rr {A,w}}  \propto M \rho_{\rr{A,w} }^{-1} R_{\rr {A,w}}^{-2}  $  [see Eq.~(\ref{eq:radius_tt})].  In the case $M=10^{17} \rr{GeV}$,  the attractor is reached inside the test mass for $\rho_{\rr A} \simeq 5 \times 10^{-20} \rr{GeV}^4$, i.e. about 1000 times lower than the aluminum density,  whereas inside the wall, it is reached for $\rho_{\rr w} \simeq 7.5 \times 10^{-20} \rr{GeV}^4 $.  This slight difference is explained by the fact that the central test mass has a diameter two times larger than the wall thickness. 

Inside the vacuum chamber, the analytical estimation is roughly recovered in the first half of the chamber.  Once in the thin-shell regime, one can also observe that the field and acceleration profiles inside the chamber are independent of the wall and mass densities,  except at their immediate vicinity. Therefore, in the deep thin-shell regime, the scalar field and acceleration both at the \textit{near} and \textit{far} positions of the interferometer do not depend on the test mass and wall densities and sizes, neither on the exterior environment. In order to obtain the numerical solution inside the chamber, down to low values of $M$, one can therefore use the trick to set the wall and mass densities high enough to be in the thin-shell regime but low enough for the field profile to be numerically tractable through the borders of the four regions. 
  
\begin{figure}
\begin{center}
\includegraphics[height=75mm, trim= 240 0 250 0,clip=true]{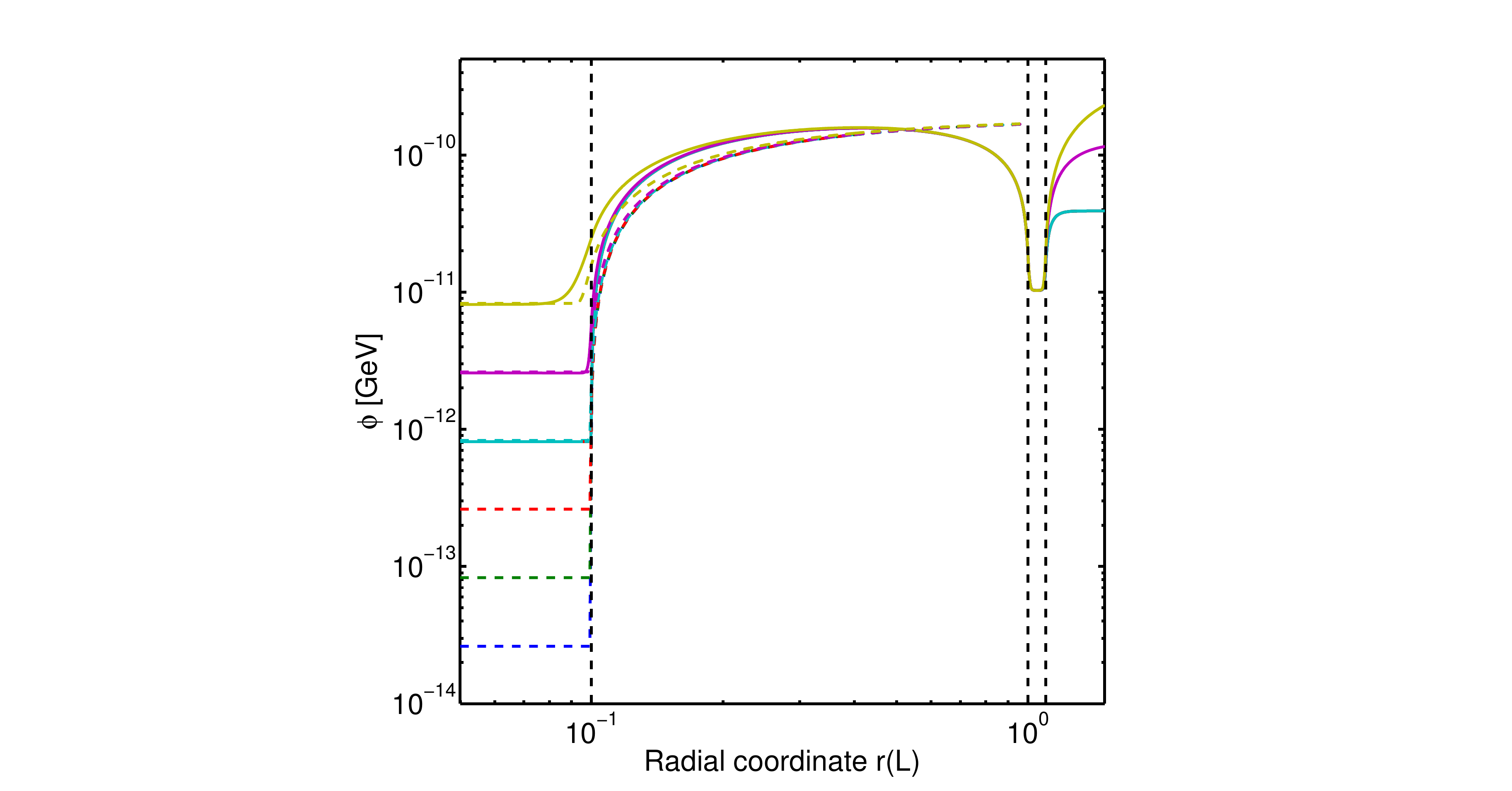}
\caption{Numerical results (solid lines) and analytical approximation (dashed lines) for the scalar field profile of the Chameleon-2 model, in the \textit{strongly perturbing} (thin shell) regime, for $\Lambda = 2.4 \times 10^{-12} \GeV$ and values of the coupling $M$ listed in Table~\ref{tab:profiles}.  The test mass, wall and exterior densities have been adapted for making the profile numerically tractable, with no effect inside the vacuum chamber (apart at the immediate vicinity of the borders), as explained in Sec.~\ref{sec:strong_field_cham_2}.  The ratios $M/ \rho$ were kept constant (with the same value than for the red curve of Fig.~\ref{plot:density_field}), which fixes the thin-shell radius, apart for $M=10^{18} \rr{GeV}$ (purple) and $M=10^{19} \rr{GeV}$ (beige) for which only the wall density was adapted.  Noticeable deviations from the analytical estimation are observed inside the chamber, due to the wall effects. 
Vertical lines mark out the four regions (test mass, chamber, wall and exterior).}
\label{plot:cham2_field}
\end{center}

\begin{center}
\includegraphics[height=75mm, trim= 240 0 250 0, clip=true]{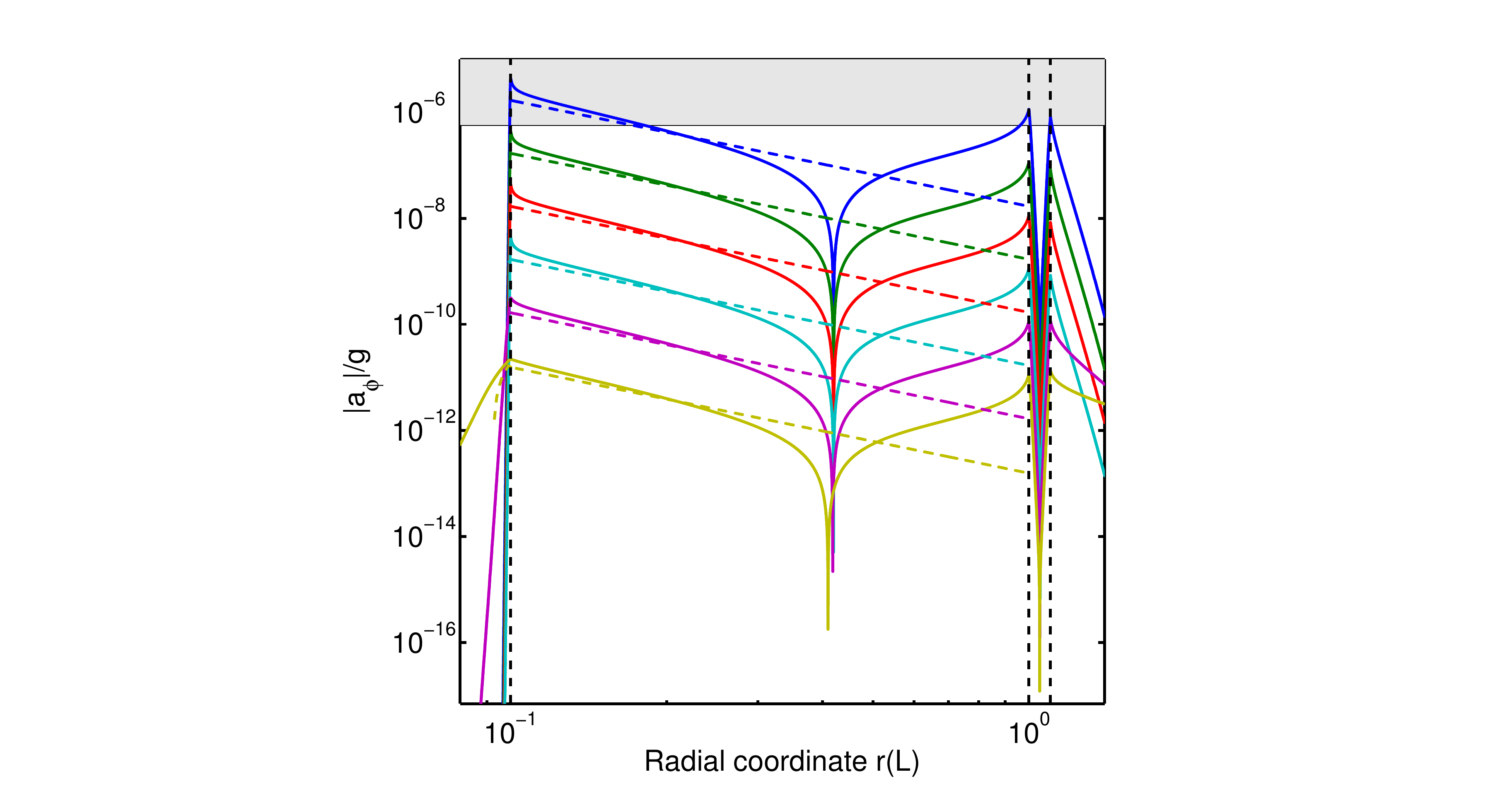}
\caption{Numerical results (solid lines) and analytical approximation (dashed lines) for the acceleration $|a_\rr\phi| / g$ profile, for the same model and parameters as in Fig.~\ref{plot:cham2_field}.  Strong discrepancies are observed between the four-region (numerical) and the two-region (analytical) models. Vertical lines mark out the four regions (test mass, chamber, wall and exterior).} 
\centering
\label{plot:cham2_acc}
\end{center}
\end{figure}

The field and acceleration profiles have been calculated numerically and compared to the analytical results, for 
several values of $M$ and $\Lambda \simeq 2.4$ meV.   These are represented on Figs.~\ref{plot:cham2_field} 
and~\ref{plot:cham2_acc}.  As expected the profiles have the same behavior than for the Chameleon-1 model (see 
Figs.~\ref{plot:profile:field_thinshell1} and~\ref{plot:profile_acc_thinshell1}).   Close to the test mass, one recovers the analytical estimation but one can nevertheless notice differences higher than 20\%. 

Close to the 
wall, the acceleration becomes negative, and its amplitude reaches values comparable to the acceleration at a position 
close to the test mass, which is a potentially measurable prediction that could be useful to discriminate between 
experimental systematic effects and an acceleration induced by the presence of some scalar field.  

In conclusion, we 
find that the 
atom-interferometry experiment of~\cite{khoury} already excludes values of the coupling parameter $M \lesssim 10^{14} 
\GeV $ at 95\% C.L.   Moreover, if the experimental sensitivity could be reduced down 
to $a_\phi / g \sim 10^{-11}$ (as it is claimed to be feasible in~\cite{Burrage}), the model would be probed nearly up to the 
Planck scale.  Finally, note that the typical field values reached inside the chamber are too low to induce large 
deviations from $A(\phi) \simeq 1$, which implies that our results are roughly independent of the power-law index 
$\alpha$. 

\subsection{Chamber geometry effects}
The numerical method used throughout this paper takes into account the effects of the chamber geometry, in the limit where 
the vacuum chamber is spherical. Exploring various chamber size and wall density, we propose to consider the possibility to 
realize the same atom interferometry experiment in a vacuum room in order to make the test of $M$ values up to the 
Planck scale possible in a near future.  The largest vacuum rooms have a radius larger than $R=10$~m and their walls
made of concrete are sufficiently large for the field to reach inside its effective potential minimum.  One can thus neglect the exterior of the 
chamber (see Sec.~\ref{sec:strong}).  The vacuum room can sustain a vacuum around $10^{-6}$ Torr (we assume 
$\rhoV= 5 \times 10^{-31} \rr{GeV}^{4}$), low enough to prevent $\phi_{\rr{bg}}$ to reach its effective potential minimum in vacuum.  

\begin{figure}
  \includegraphics[scale=0.38, trim=220 0 240 0, clip=true]{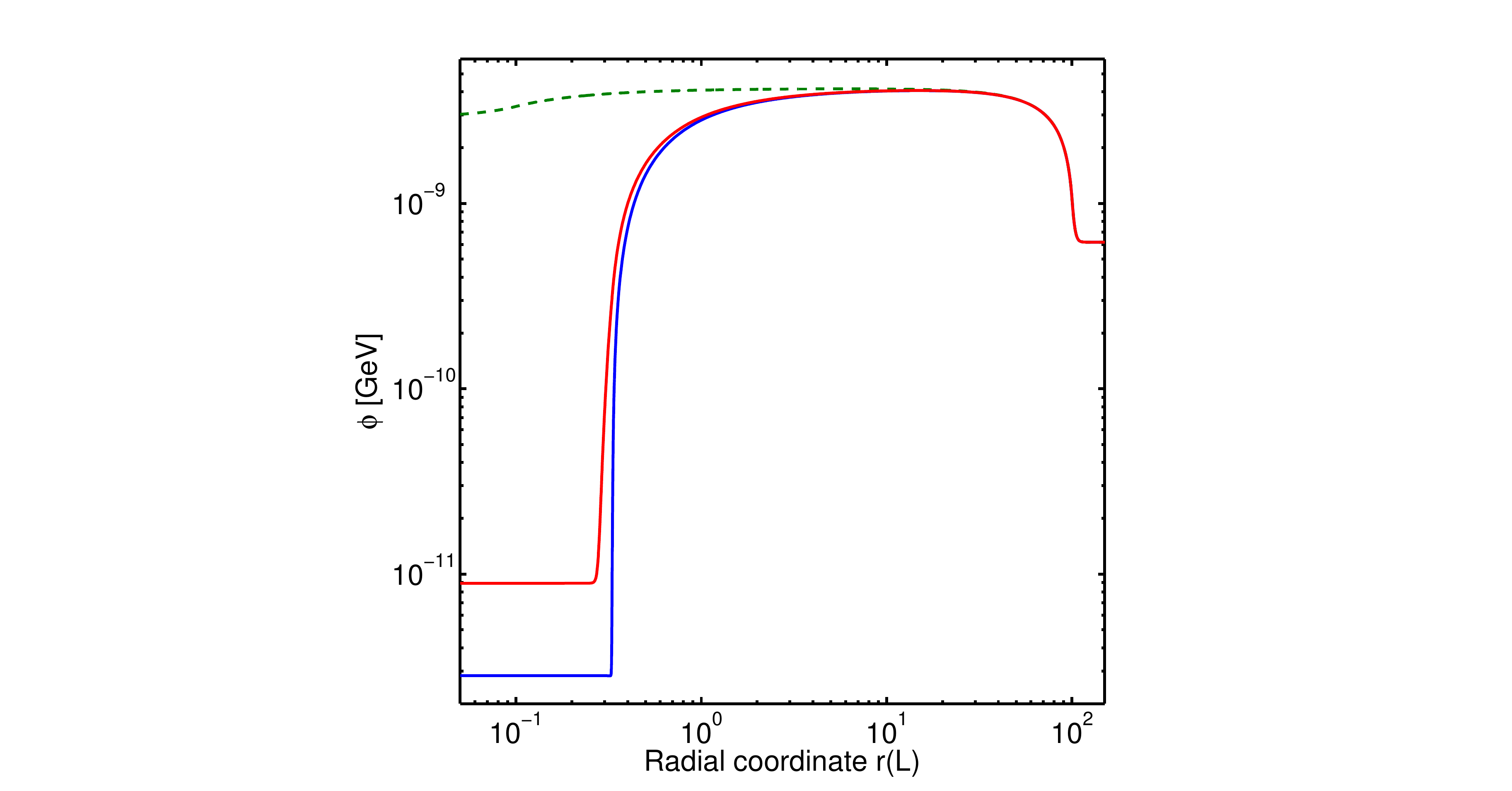}
  \caption{Numerical scalar field profiles for a vacuum room ($L=10$~m). The green dashed curve is obtained for a test 
   mass of $\RA=$1~cm with $M=m_\rr{p}$ ($\rhoW=2.5\times10^{-21}\,\rr{GeV}^4$) while the blue and the red ones 
   are obtained for $\RA=3.3$~cm with $M=0.1\times m_\rr{p}$ ($\rhoW=2.5\times10^{-22}\,\rr{GeV}^4$) and $M=m_\rr{p}$   
   ($\rhoW=2.5\times10^{-21}\,\rr{GeV}^4$) respectively. We only consider a three regions model, neglecting the effect 
of the exterior of the vacuum room (see discussion in Sec.~\ref{sec:strong}).}
  \label{plot:vac_room_field}
\end{figure}

\begin{figure}
  \includegraphics[scale=0.38, trim=220 0 240 0, clip=true]{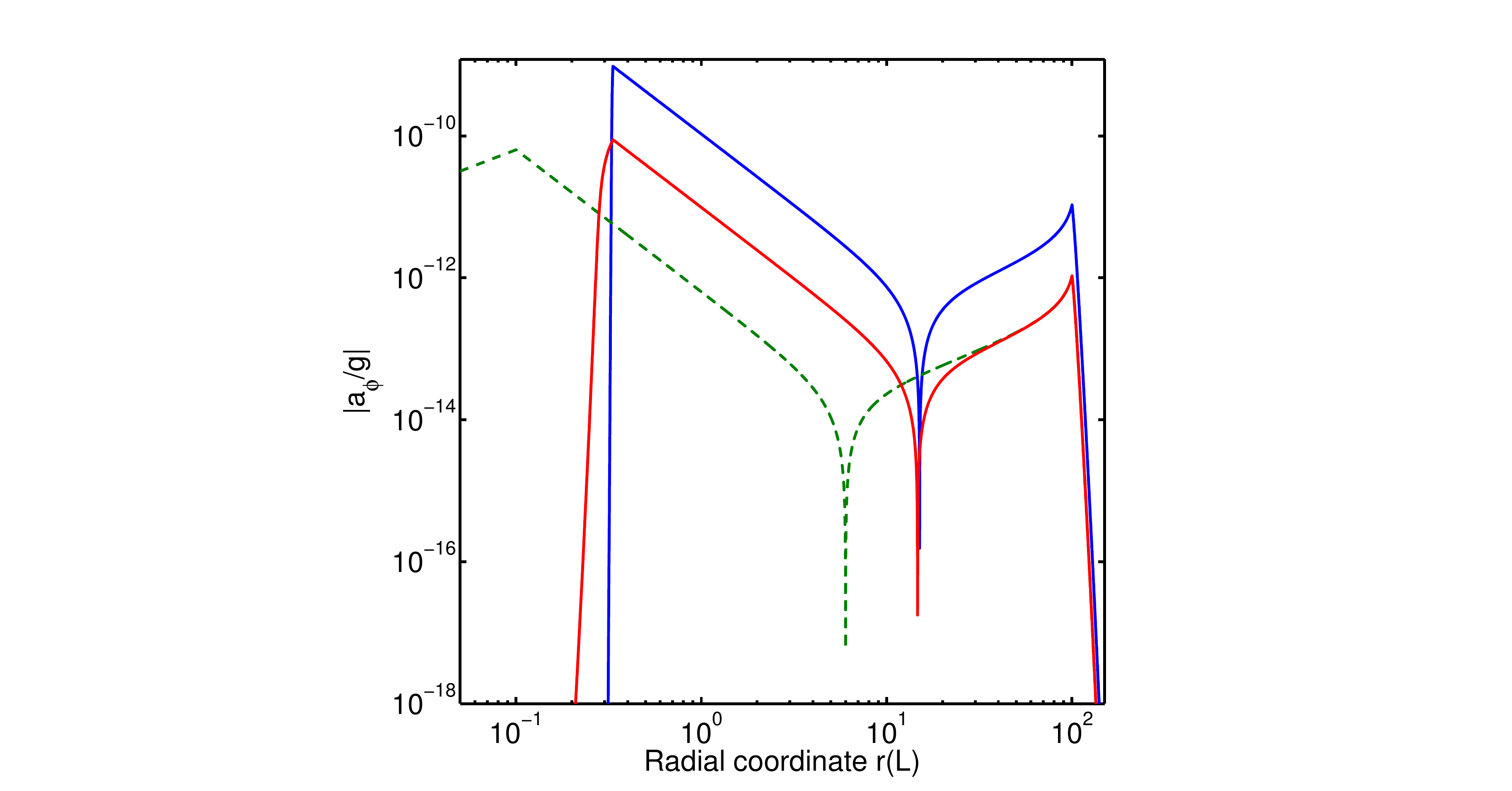}
   \caption{Numerical acceleration profiles $|a_\phi/g|$ for a vacuum room ($L=10$~m). The green dashed curve is obtained for a test 
   mass of $\RA=$1~cm with $M=m_\rr{p}$ ($\rhoW=2.5\times10^{-21}\,\rr{GeV}^4$) while the blue and the red ones 
   are obtained for $\RA=3.3$~cm with $M=0.1\times m_\rr{p}$ ($\rhoW=2.5\times10^{-22}\,\rr{GeV}^4$) and $M=m_\rr{p}$   
   ($\rhoW=2.5\times10^{-21}\,\rr{GeV}^4$) respectively. We only consider a three regions model, neglecting the effect 
of the exterior of the vacuum room (see discussion in Sec.~\ref{sec:strong}).}
  \label{plot:vac_room_acc}
\end{figure}

Numerical field and acceleration profiles are reported on Figs.~\ref{plot:vac_room_field} and \ref{plot:vac_room_acc} 
respectively. Assuming as before $\rhoA=1.2\times 10^{-17}~\rr{GeV}^4$, it results that 
a test mass of 1~cm radius only enables to probe the regime where the field does not reach $\phi_{\rr A}$ inside the test mass (see dashed green lines on 
Figs.~\ref{plot:vac_room_field} and \ref{plot:vac_room_acc}), the acceleration being thus poorly constrained.  However, 
provided that the test mass radius is larger (e.g. $\RA=3.3$~cm), the strongly perturbing regime is reached and the 
acceleration is large enough to be measurable in a near future for $M$ of the order 
of $m_\rr{pl}$. As a result, for $M=m_\rr{pl}$, $|a_\phi|/g=2.4\times 10^{-10}$ at 8.8~mm from the 
surface of the test mass (the previously called \textit{near} position in Sec.~\ref{sec3}) while $|a_\phi|/g=5.7\times 
10^{-10}$ for $M=0.1~m_\rr{pl}$. In comparison, the test mass of 1~cm gives rise to $|a_\phi|/g=1.7\times 10^{-11}$ for $M=m_\rr{pl}$. 

Similarly to what was obtained in Sec.~\ref{sec:strong}, the thin shell regime cannot be tracked numerically if the wall density is of the order of the concrete $\rho\sim 10^{-17}\,\rr{GeV}^4$.  But one can safely consider lower values of 
$\rhoW$ (see Fig.~\ref{plot:vac_room_field}) without any significant change of the results inside the vacuum room.

\section{Conclusion} \label{sec_CCL}

The chameleon screening mechanism is able to suppress the fifth-force induced by a scalar field in locally dense environment, 
while allowing the scalar field to be responsible for dark energy on large astrophysical scales and thus to affect significantly 
the large scale structure formation.   This makes chameleon theories good candidates for explaining dark energy 
and for being testable by future cosmology-dedicated experiments, such as Euclid~\cite{Amendola:2012ys} 
or the next generation of giant radio-telescopes dedicated to 21~cm cosmology~\cite{Brax:2012cr}.  
Chameleon theories are also well constrained by local tests of gravity in the solar system, in the galaxy, 
as well as in laboratory.   Recently it has been proposed to use an atom-interferometry experiment 
to constrain chameleon models with an unprecedented accuracy by probing the acceleration induced by the presence of the scalar field on cold atoms.  The experiment is realized inside a vacuum chamber in order to reduce the screening effect, and a central mass is used to source some field gradient.   Forecasts were calculated in~\cite{Burrage} and a first experimental setup was build and used to establish new constraints on chameleon model~\cite{khoury}.   However the calculations of the field and acceleration profiles rely on several approximations, and until now did not fully consider the effects of the vacuum chamber wall and of the exterior environment.

The purpose of this work was to validate and refine previous calculations, by using a numerical approach consisting in solving the Klein-Gordon equation in the static and spherically symmetric case for a four-region model representing the central source mass, the vacuum chamber, its wall, and the exterior environment.   Three boundary conditions are imposed:  the field must be regular at origin and reach the minimum of the effective potential with a Yukawa profile, at large distance in the exterior environment.  Our method allows to probe the transition between the regime where the central source mass only weakly perturbs the field configuration, and the thin-shell regime where the field inside the central mass and inside the chamber walls reaches the minimum of the effective potential over a very small distance.   Two typical chameleon potentials were considered, in inverse power-laws and allowing varying powers, as well as a standard exponential form for the coupling function.  

In the weakly perturbing regime, it is found that the chamber wall enhances significantly the scalar field inside the vacuum chamber and reduces the induced acceleration, by up to one order of magnitude compared to previous analytical estimations and with a maximal effect close to the wall. 

Going to the thin-shell regime, for our fiducial experimental setup, the field reaches the 
attractor inside the chamber wall and the exterior environment becomes thus irrelevant.  However, for reasonable value of the 
induced acceleration, the field inside the vacuum chamber does not reach the minimum of the effective potential and is 
instead related to the size of the chamber, as first noticed in~\cite{Burrage}.  Our analysis refines the field and acceleration profiles in the chamber and highlights noticeable deviations 
from the analytical estimation, which is nevertheless roughly recovered close to the central test mass.  
Close to the chamber wall, the 
acceleration becomes negative, with a magnitude similar to the one close to the central mass.  We argue that this 
prediction could be useful to distinguish between systematic effects and fifth-force effects which should be 
\textit{maximal and opposite close to the central mass and to the wall}, and should vanish roughly at the 
middle-distance between the test mass and the walls.   

Refined constraints have been derived on the coupling parameter $M$ from the atom-interferometry experiment of~\cite{khoury}.  For the chameleon potential $V(\phi) = \Lambda^{4+\alpha} / \phi^\alpha$ and a coupling function $A(\phi) = \exp(\phi /M)$, one finds $M \gtrsim 7 \times 10^{16} \GeV$, independently of the power-law.   For the bare potential $V(\phi) = \Lambda^4 (1+ \Lambda/ \phi)$, we find that 
$M \gtrsim  10^{14} \GeV$.   
We have also confirmed that a future experiment reducing its sensitivity down to $a \sim 10^{-10} \rr{m/s^2}$ would be able to rule out most of the parameter space of the latter model, nearly up to the Planck scale. 

Finally, we have proposed to realize a similar atom-interferometry experiment inside a vacuum room.  The density inside such rooms is low enough for the field profile and the induced acceleration to depend only on the size of the room.  If the room radius is larger than about $10$ meters, we find that the chameleon model could be probed up to the Planck scale.  Nevertheless, further work is needed to implement realistic non-spherical geometries of the room (or of the vacuum chamber).  

We conclude that numerical results will be helpful in the future in order to establish accurate bounds on various modified gravity models. In particular, the effects of the vacuum chamber wall and its exterior environment cannot be neglected. Our numerical method is easily extendable to study other forms of the field potential and other modified gravity models requiring a screening mechanism, such as the symmetron, dilaton and f(R) models. An investigation of the symmetron model is in progress and should be released soon. Finally it can be easily adapted to other experiments.

\section{Acknowledgments}

We warmly thank Christophe Ringeval, Clare Burrage, Holger M\"uller, Justin Khoury, Benjamin Elder and Philipp Haslinger for useful comments and discussion.  The work of S.C. is partially supported by the \textit{Return Grant} program of the Belgian Science Policy (BELSPO).   S.S. is supported by the FNRS-FRIA and A. F. is partially supported by the ARC convention No. 11/15-040.   

\bibliography{biblio}

\end{document}